\documentclass[aps,preprint,prd,nofootinbib]{revtex4}

\usepackage[dvipsnames]{xcolor}
\usepackage{graphics,tikz,tikzscale,float}
\usepackage{ulem}
\usepackage[caption = false]{subfig}

\usepackage{graphicx}
\usepackage{epsfig}
\usepackage{epstopdf}
\usepackage[utf8]{inputenc}
\usepackage{amsmath}
\usepackage{amsfonts}
\usepackage{amssymb}
\usepackage{slashed}
\usepackage{booktabs}
\usepackage{multirow}
\usepackage{txfonts}
\allowdisplaybreaks

\newcommand{\V}{\mathit{V_N}}
\newcommand{\Ve}{\mathit{V_E}}
\newcommand{\Vk}{\mathit{V_K}}
\newcommand{\Vek}{\mathit{V_{EK}}}

\newcommand{\sle}{\ensuremath{\slashed\epsilon}}

\newcommand{\slk}{\ensuremath{\slashed k}}
\newcommand{\nn}{\nonumber}

\begin{document}

\title{Pion photoproduction off nucleons in covariant chiral perturbation theory}

\author{Gustavo H. Guerrero Navarro}\email{gusguena@ific.uv.es}
\affiliation{Departamento de F\'{\i}sica Teorica and Instituto de Fisica Corpuscular (IFIC),
Centro Mixto UVEG-CSIC, Valencia E-46071, Spain}

\author{Astrid N.~Hiller Blin}\email{hillerbl@uni-mainz.de}
\affiliation{Institut f\"ur Kernphysik \& PRISMA$^+$ Cluster of Excellence, Johannes Gutenberg Universit\"at, D-55099 Mainz, Germany}

\author{M. J. Vicente Vacas}\email{vicente@ific.uv.es}
\affiliation{Departamento de F\'{\i}sica Teorica and Instituto de Fisica Corpuscular (IFIC),
Centro Mixto UVEG-CSIC, Valencia E-46071, Spain}

\author{De-Liang Yao}\email{yaodeliang@hnu.edu.cn}
\affiliation{School of Physics and Electronics, Hunan University, Changsha 410082, China}
\date{\today}

\begin{abstract}
Pion photoproduction off the nucleon close to threshold is  studied in covariant baryon chiral perturbation theory at O($p^3$)  in the extended-on-mass-shell scheme, with the explicit inclusion of the $\Delta(1232)$ resonance using the $\delta$ counting. The theory is  compared to the available data of cross sections and polarization observables for all the charge channels. Most of the necessary low energy constants are well known from the analysis of other processes and the comparison with data  constrains some of the still unknown ones.  The $\Delta(1232)$ contribution is significant in improving the agreement with data, even at the  low energies considered. 
 
\end{abstract}

\maketitle

\section{Introduction}

Single pion photoproduction off the nucleons has been a subject of  strong and continuous theoretical and experimental efforts. Many have been dedicated to the investigation of the process at intermediate energies which allowed us to study the spectrum and properties of numerous baryon resonances~\cite{Adler:1968tw,Drechsel:2007if}. Here, we address the near threshold region, where  chiral perturbation theory (ChPT)~\cite{Weinberg:1978kz,Gasser:1983yg,Gasser:1984gg,Scherer:2012xha}, the effective field theory of QCD at low energies, should provide an adequate framework and only nucleons, pions, and the lowest lying resonances might play a role.

Early work attempted to describe this process  through low-energy theorems (LET) obtained from gauge and Lorentz invariance~\cite{PhysRev.93.233} and later from  current algebra and the partial conservation of the axial-current~\cite{DeBaenst:1971hp,Vainshtein:1972ih}. These theorems described well the production of charged pions, but failed for the case of the $\gamma + p \rightarrow p + \pi^0$ process~\cite{Mazzucato:1986dz,Beck:1990da,Drechsel:1992pn,Bernard:2006gx}.
In one of the earliest successes of ChPT with baryons, Bernard {\it et al.}~\cite{Bernard:1991rt,Bernard:1992nc,Bernard:2001gz} could solve the discrepancies between theory and the data available at the time with corrections related to loop-diagram contributions. Still, the theoretical models showed their limitations with the more precise measurements   of cross sections and polarization observables  obtained at the Mainz Microtron (MAMI) in 2013~\cite{Hornidge:2012ca}. 
For instance, it was found in an $\mathcal{O}(p^4)$ heavy baryon (HB) calculation that the agreement with data was satisfactory only up to some 20~MeV above threshold~\cite{FernandezRamirez:2012nw}.
This indicated the need of calculations at even higher orders.
The situation was not better in other  approaches to baryon ChPT, such as the extended-on-mass-shell (EOMS) scheme, which at $\mathcal{O}(p^4)$ also obtained a good agreement  only for a very limited range of energies~\cite{Hilt:2013uf}. 
 However, a higher order calculation, apart from the added technical complications, would  lose its predictive power because of the new set of undetermined low energy contants (LECs) appearing in the Lagrangian. These difficulties could be intrinsic to this specific process, for instance because of the cancellation happening at the lowest order of the chiral expansion, but they may also signal the need for some revision of the theoretical approach. 

In recent years, there has been a considerable advance in the qualitative and quantitative understanding of low energy hadron physics using  ChPT.  It provides a systematic framework to obtain a perturbative expansion in terms of small meson masses and external momenta and has a quite impressive record on its predictivity and the quality of its description of multiple observables involving mesons, nucleons and photons~\cite{Bijnens:2014lea,Bernard:2007zu}. Nonetheless, ChPT, when applied to systems with baryons, contains some subtleties which may hinder its progress.
As  shown in Ref.~\cite{Gasser:1987rb}, in the presence of baryon loops the naive power counting is broken because of the nonzero nucleon mass in the chiral limit, making difficult the development of a scheme that allows for a systematic evaluation of higher orders in the chiral expansion. This problem was first solved in the HB formalism (HBChPT)~\cite{Jenkins:1990jv,Jenkins:1991es}, at the expense of  losing Lorentz covariance, and later by some covariant methods  as the infrared regularization (IrChPT)~\cite{Becher:1999he} and the EOMS scheme~\cite{Fuchs:2003qc}~\footnote{See, e.g., Ref.~\cite{Scherer:2012xha} for a review of the three schemes and the discussion  in the introduction of Ref.~\cite{Alarcon:2012kn}.}. Here, we  adhere to the latter, which, apart from providing a proper power counting, preserves the analytic structure of the calculated amplitudes. Furthermore, it usually leads to a faster chiral convergence than HBChPT or IrChPT~\cite{Alarcon:2012kn,Geng:2008mf,MartinCamalich:2010fp}.  This approach has  been used with satisfactory results in the calculation of many  baryon observables  such as masses, magnetic moments, axial form factors, among others~\cite{Fuchs:2003ir,Lehnhart:2004vi,Schindler:2006it,Schindler:2006ha,Geng:2008mf,Geng:2009ik, MartinCamalich:2010fp,Alarcon:2011zs,Ledwig:2011cx,Chen:2012nx,Alvarez-Ruso:2013fza,Ledwig:2014rfa,Lensky:2014dda}.
Moreover, it has been successfully applied to many processes, among which $\pi N$ scattering~\cite{Alarcon:2012kn,Chen:2012nx,Yao:2016vbz,Siemens:2016hdi} and the pion electromagnetic production on the nucleons~\cite{Hilt:2013uf,Blin:2014rpa,Blin:2016itn,Hilt:2013fda}. Both are directly related to the pion photoproduction on the nucleons  investigated in this work. 
 
In addition to the power counting problem in baryon ChPT, another issue arises  due to  the small mass difference between the nucleon and the  $\Delta(1232)$ resonance. In fact,  the mass of the latter is little  above the pion production threshold. Due to this proximity of the resonance to the threshold and its large transition  couplings to pions and photons,  the $\Delta(1232)$ is crucial for the description of $\pi N$ and $\gamma N$ processes even at very low energies~\cite{Ericson:1988gk}.  These facts suggest the  importance of the explicit inclusion of the $\Delta(1232)$ resonance in our effective theory. The hope is that the incorporation of the most relevant degrees of freedom, such as those associated to the $\Delta$, could lead to a faster convergence of the chiral series. A price to pay is the emergence of a new small parameter, $\delta=m_\Delta-m_N\approx 300 $~MeV that should be properly accounted for in the chiral expansion, where $m_\Delta$ and $m_N$ are the masses of the $\Delta$ resonance and the nucleon, respectively. 
 
In this work, we investigate  the near threshold pion photoproduction off nucleons within the aforementioned effective theory approach, i.e. EOMS ChPT,  at $O(p^3)$.  We also choose to include the $\Delta(1232)$ resonance explicitly. This, or a very similar approach has already been used for the analysis of Compton scattering~\cite{Lensky:2009uv,Blin:2015era}, $\pi N$ scattering~\cite{Alarcon:2012kn,Yao:2016vbz,Siemens:2016hdi} or the weak process $\nu N\rightarrow l N'\pi $~\cite{Yao:2018pzc,Yao:2019avf} of  high interest for neutrino detection.   Moreover, the fact that we are using the same framework  at the same chiral order  as some of these works allows us to fix many of the LECs of the theoretical model. 

Besides the general reasons given in the previous paragraphs, the inspection of the cross section data shows that the $\Delta(1232)$ resonance is conspicuously dominant for all  pion photoproduction channels~\cite{Ericson:1988gk}, and its tail could well be large even close to threshold.  The possible importance of the $\Delta$ resonance for the chiral analyses of pion photoproduction had already been suggested in Refs.~\cite{Hemmert:1996xg,Bernard:2001gz,Hornidge:2012ca,FernandezRamirez:2012nw}. Indeed, the convergence of the chiral series and the agreement with data was found to improve substantially with this inclusion in the investigation of the neutral pion photoproduction on the proton in Refs.~\cite{Blin:2014rpa,Blin:2016itn}. Nevertheless, it is also clear from even a cursory perusal of data, that the energy dependence is very different for the $\gamma p\rightarrow \pi^0p$ and the $\gamma p\rightarrow \pi^+n$ processes. In the latter case there is a large nonresonant electric dipole contribution that produces $s$-wave pions and is strongly suppressed in the $\pi^0$ case. For this reason, the importance of including higher orders and the $\Delta$ resonance is especially strong for the neutral pion channel. However, also the other ones will have noticeable corrections due to these inclusions. Furthermore, the different channels of pion production are sensitive to different ChPT LECs, leading to the need of studying carefully also the charged pion channels in the same framework.

Motivated by the sensitivity to different mechanisms of the various channels, here we extend the analysis of Refs.~\cite{Blin:2014rpa,Blin:2016itn}, restricted to the   $\gamma p\rightarrow \pi^0p$ process, by incorporating the other channels, in which charged pions are produced. We  perform a global study of all the data currently available in the low energy region. This  amounts to measurements of angular distributions, total cross sections and spin observables such as beam and target asymmetries.  Ultimately, these studies will benchmark the ability to improve upon the predictions for the weak pion production processes~\cite{Bernard:1993xh,Alvarez-Ruso:2017oui,Yao:2018pzc,Yao:2019avf}, for which the data are very scarce, and integrated over wide ranges of energies, thus making it impossible to constrain well the LECs or to make concrete statements about the behavior at specific energies. While the predictive power of ChPT calculations is limited to the threshold region, they should properly be taken into account in phenomenological models that aim to describe weak pion production in wider energy regions, see also Ref.~\cite{Alvarez-Ruso:2014bla} and references therein.

The inclusion of the charged channels requires the addition of a more extensive set of diagram topologies and also of some extra pieces of the chiral Lagrangian with their corresponding LECs. Furthermore, we  incorporate a more detailed analysis of the errors, estimating both the statistical uncertainty coming from the fits and the uncertainty related to the truncation of the chiral series.

The structure of the paper is as follows. In Sec.~\ref{sec:form}, we present the basic formalism, the chiral Lagrangian and the theoretical model for the amplitude. Sec.~\ref{sec:fiterr} describes the experimental database and the fit method, including the procedure for the error estimation. Finally, results are presented in Sec.~\ref{sec:res}. We summarize in Sec.~\ref{sec:summary}. 

\section{Basic formalism and theoretical model}
\label{sec:form}
\begin{figure}
\begin{center}
\includegraphics[width=0.3\textwidth]{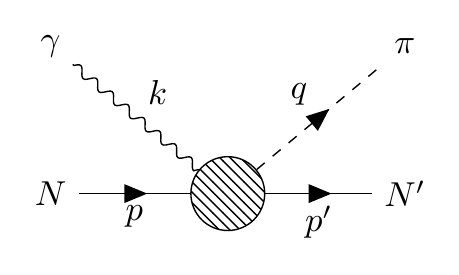}
\caption{Kinematics of the pion photoproduction process. $k$ and $p$ are the incoming photon and nucleon momenta, while $q$ and $p'$ correspond to the outgoing pion and nucleon momenta, respectively.}
\label{fg:feynmanBlob}
\end{center}
\end{figure}

\subsection{Kinematics, amplitude decomposition and observables }
\label{subsec:kin.amp.obs}

The pion photoproduction off the nucleon, depicted in Fig.~\ref{fg:feynmanBlob}, can occur in four possible charge channels: $    \gamma p  \rightarrow \pi^0 p, 
    \gamma p  \rightarrow \pi^+ n, 
    \gamma n  \rightarrow \pi^- p,
    \gamma n  \rightarrow \pi^0 n$. 
The differential cross section in the center of mass (c.m.) system can be written as
\begin{equation}\label{eq:dsigmado}
\frac{d \sigma}{d \Omega_\pi}= \frac{1}{64 \pi^2 s}\frac{\Lambda^{1/2}(s,m_N^2,m_\pi^2)}
{\Lambda^{1/2}(s,m_N^2,0)} \frac{1}{4} \sum_{s_i,s_f,\lambda} \left| \mathcal{T} \right|^2, 
\end{equation}
where $s\equiv(k+p)^2$ is the Mandelstam variable, $\Lambda(x,y,z)=(x-y-z)^2-4yz$ is the Källén function, $m_N$ and $m_\pi$ are the nucleon and pion masses, respectively. The modulus squared of the scattering amplitude $\mathcal{T}$ is averaged over the initial nucleon spin ($s_i$) and photon polarization ($\lambda$) and summed over the final nucleon spin ($s_f$).
For practical purposes, it is convenient to use a representation of $\mathcal{T}$ in terms of the Chew-Goldberger-Low-Nambu (CGLN) amplitudes $\mathcal{F}_i$~\cite{Chew:1957tf}, which lead to simple expressions for multipoles, cross sections and the polarization observables.  In the CGLN formalism, $\mathcal{T}$ can be written as
\begin{align}
\mathcal{T} = \frac{4\pi \sqrt{s}}{m_N}\chi_f^\dagger\mathcal{F}\chi_i,
\label{eq:scatt.amp}
\end{align}
where $\chi_i$ and $\chi_f$ are Pauli spinors of the initial and final nucleon states, respectively.
For real photons and in the Coulomb gauge ($\epsilon^0=0$, $\vec{\epsilon}\cdot\vec{k}=0$), the amplitude $\mathcal{F}$ may be decomposed as
\begin{align}
\mathcal{F} = \mathrm{i}\,\vec\sigma\cdot\vec\epsilon\,\mathcal{F}_1
+  \vec\sigma\cdot\hat q\,\vec\sigma\cdot\hat k\times\vec\epsilon\,\mathcal{F}_2
+  \mathrm{i}\,\vec\sigma\cdot\hat k\, \hat q\cdot \vec{\epsilon}\,\mathcal{F}_3
+  \mathrm{i}\,\vec\sigma\cdot\hat q \,\hat q\cdot \vec{\epsilon}\,\mathcal{F}_4,
\label{eq:Fdecomposition}
\end{align}
with $\vec\sigma$ the Pauli matrices, $\epsilon$ the photon polarization and $\hat q,\,\hat k$ unit vectors in the direction of $\vec q$ and $\vec k$, respectively. The explicit expressions for each $\mathcal{F}_i$ amplitude are given in Appendix \ref{app:Amp.rep},   Eqs.~(\ref{coefs:F1})-(\ref{coefs:F4}). In this representation, the unpolarized angular cross section in the c.m. system in Eq.~\eqref{eq:dsigmado} is recast as
\begin{align}
\label{eq:dsigma}
 \frac{d \sigma}{d \Omega_\pi}=d\sigma_0 &= \rho_0\,\mathfrak{R}e \left\lbrace \mathcal{F}^*_1 \mathcal{F}_1 + \mathcal{F}^*_2 \mathcal{F}_2 + \sin^2 \theta \,(\mathcal{F}^*_3 \mathcal{F}_3 + \mathcal{F}^*_4 \mathcal{F}_4)/2 \right. \nn \\
& \left. + \sin^2 \theta\, (\mathcal{F}^*_2 \mathcal{F}_3 + \mathcal{F}^*_1 \mathcal{F}_4 + \cos \theta\, \mathcal{F}^*_3 \mathcal{F}_4) - 2 \cos \theta \,\mathcal{F}^*_1 \mathcal{F}_2 \right\rbrace , 
\end{align}
where $\theta$ stands for the scattering angle  between the incoming photon and the outgoing pion and
\begin{equation}
\rho_0 = \frac{\Lambda^{1/2}(s,m_N^2,m_\pi^2)}
{\Lambda^{1/2}(s,m_N^2,0)}=\frac{|\vec {q}|}{|\vec {k}|},
\end{equation}
with $|\vec{q}|$ and $|\vec{k}|$ evaluated in the c.m. system.

At the studied energies, apart from the unpolarized angular cross section, there are many data for the polarized photon asymmetry. This  observable is defined by 
\begin{equation}
\Sigma\equiv\frac{\mathrm{d}\sigma_\perp-\mathrm{d}\sigma_\parallel}{\mathrm{d}\sigma_\perp+\mathrm{d}\sigma_\parallel},
\end{equation}
with $d\sigma_\perp$ and $d\sigma_\parallel$ the angular cross sections for photon polarizations perpendicular and parallel to the reaction plane, respectively. In the CGLN representation we have~\cite{Sandorfi:2010uv}
\begin{equation}\label{eq:Sigma}
\Sigma= -\frac{\rho_0}{d\sigma_0}\,\sin^2 \theta \; \mathfrak{R}e \left\lbrace (\mathcal{F}^*_3 \mathcal{F}_3 + \mathcal{F}^*_4 \mathcal{F}_4)/2 + \mathcal{F}^*_2 \mathcal{F}_3 + \mathcal{F}^*_1 \mathcal{F}_4 + \cos \theta \,\mathcal{F}^*_3 \mathcal{F}_4 \right\rbrace.
\end{equation}
  In its turn, the target asymmetry, defined as the ratio 
\begin{equation}
T=\frac{\mathrm{d}\sigma_+-\mathrm{d}\sigma_-}{\mathrm{d}\sigma_++\mathrm{d}\sigma_-},
\end{equation}
where $d\sigma_+$ and $d\sigma_-$ correspond to the cross sections for target nucleons polarized up and down in the direction of $\vec{k}\times \vec{q}$, can be written as
\begin{equation}\label{eq:Tasym}
T= \frac{\rho_0}{d\sigma_0} \sin \theta\; \mathfrak{I}m \left\lbrace \mathcal{F}^*_1 \mathcal{F}_3 - \mathcal{F}^*_2 \mathcal{F}_4 + \cos \theta (\mathcal{F}^*_1 \mathcal{F}_4 - \mathcal{F}^*_2 \mathcal{F}_3) - \sin^2 \theta \mathcal{F}^*_3 \mathcal{F}_4 \right\rbrace . 
\end{equation}
Useful expressions for other polarization observables in terms of the $\mathcal{F}_i$ amplitudes can be found, for instance, in Ref.~\cite{Sandorfi:2010uv}.

\subsection{Power counting and chiral Lagrangians}

As was discussed in the Introduction, when the $\Delta(1232)$ resonance is  explicitly included   a new small parameter $\delta=m_{\Delta}-m_N\approx 300$~MeV  appears, which must be taken into account in the chiral expansion. In this work, we use the $\delta$ counting, introduced in Ref.~\cite{Pascalutsa:2002pi}, in which $\delta\sim O(p^{1/2})$. Therefore, the chiral order $D$ of a diagram with $L$ loops, $V^{(k)}$ vertices of $O(p^{k})$, $N_\pi$ internal pions, $N_N$ nucleon propagators and $ N_\Delta$ $\Delta(1232)$ propagators is given by
\begin{align}
D=4L+\sum_{k=1}^{\infty}{kV^{(k)}}-2N_\pi-N_N-\frac12 N_\Delta.
\end{align}
Here, we  consider all contributions up through $O(p^{3})$. The following pieces of the chiral effective Lagrangian are required,
\begin{align}
\mathcal{L}_{\rm eff}=\sum_{i=1}^2\mathcal{L}_{\pi\pi}^{(2i)}+\sum_{j=1}^3\mathcal{L}_{N}^{(j)}+{\cal L}^{(1)}_{\pi N\Delta}+{\cal L}^{(2)}_{\gamma N\Delta}\ ,
\end{align}
where the superscripts represent the chiral order of each of the terms. The needed terms of the pionic interaction are given by \cite{Gasser:1983yg,Gasser:1987rb}
\begin{eqnarray}
\mathcal{L}_{\pi\pi}^{(2)} &=& \frac{F_0^2}{4} \text{Tr} \left[ \nabla^\mu U \left( \nabla_\mu U \right)^\dagger  + \chi U^\dagger + U \chi^\dagger \right],
\label{eq:LOLagpipi}\\
\mathcal{L}_{\pi\pi}^{(4)}&=&\frac{\ell_4}{16}\left (({\rm Tr}[\chi U^\dagger+U\chi^\dagger])^2+2{\rm Tr}[\nabla_\mu U(\nabla^\mu U)^\dagger]{\rm Tr}[\chi U^\dagger+U\chi^\dagger]\right)+\cdots\,,\label{eq:LagPi4}
\end{eqnarray}
where the Goldstone pion fields are written in the isospin decomposition

\begin{equation}
U=\exp \left[ i \frac{\vec{\tau} \cdot \vec{\pi}}{F_0} \right]=
\exp
\left[ \frac{i}{ F_0}
\left(
\begin{array}{cc}
   \pi^0  & \sqrt{2}\,\pi^+ \\
 \sqrt{2}\,\pi^-   & -\pi^0
\end{array}
\right)\right]\,,
\end{equation}
$\tau^i$ ($i=1,2,3$) are the Pauli matrices, $F_0$ indicates the pion decay constant in the chiral limit, and $\nabla_\mu U = \partial_\mu U - i r_\mu U + i U l_\mu$  is the covariant derivative with external fields $r_\mu=l_\mu = e Q A_\mu$. Here, $e$ is  the electric charge of the electron, $Q=(\tau^{3} + \mathbf{1}_{2 \times 2})/2$  is the charge matrix, and $A_\mu$ is the photon field. ${\rm Tr}[\cdots]$ denotes the trace in flavor space. We will be working in the isospin symmetric limit, and thus $\chi=m_{\pi}^2 \,\mathbf{1}_{2 \times 2}$, with $m_\pi$ the corresponding pion mass.

The relevant terms that describe the interaction with nucleons at $O(p^1)$ are given by \cite{Fettes:2000gb}

\begin{equation}
\mathcal{L}_{N}^{(1)} = \bar{N} \left( i \slashed D - m + \frac{g}{2} \slashed u \gamma_5 \right) N,
\end{equation}
where $N=(p,n)^T$ is the nucleon doublet with mass $m$ and axial charge $g$, both in the chiral limit. Furthermore,
\begin{equation}
D_\mu=\partial_\mu + \Gamma_\mu,\quad
\Gamma_\mu = \frac{1}{2} [u^\dagger, \partial_\mu u] - \frac{i}{2} u^\dagger r_\mu u - \frac{i}{2} u l_\mu u^\dagger, \quad
u=U^{1/2},\quad
 u_\mu=i u^\dagger (\nabla_\mu U) u^\dagger.
\end{equation}
 At the second order the only relevant terms are
\begin{equation}
\mathcal{L}_{N}^{(2)}  = \bar{N} \left( c_1 \text{Tr}\left[ \chi_{+} \right]+  \frac{c_6}{8m} F_{\mu\nu}^{+}\sigma^{\mu\nu}  + \frac{c_7}{8m} \text{Tr} \left[ F_{\mu\nu}^{+} \right] \sigma^{\mu\nu} \right) N +\cdots,
\label{LN2}
\end{equation}
with $\chi_{\pm} = m_\pi^2 \left( U^\dagger \pm U \right)$ in the isospin limit, $F^{\pm}_{\mu \nu} = u^\dagger F_{R \mu \nu} u \pm u F_{L \mu \nu} u^\dagger$, where in our case $F_{R \mu \nu}=F_{L \mu \nu}=F_{ \mu \nu}$ results in the electromagnetic tensor $F_{ \mu \nu}=e Q (\partial_\mu A_\nu - \partial_\nu A_\mu)$. Here, $c_i$ ($i=1,6,7$) are $O(p^2)$ LECs in units of ${\rm GeV}^{-1}$.

The contributing terms of $O(p^3)$ are~\cite{Fettes:2000gb}
\begin{align}
\mathcal{L}_{N}^{(3)} &= d_8 \bar{N} \left[ \frac{1}{2m}i\epsilon
^{\mu\nu\alpha\beta}\text{Tr} \left[ \widetilde{F}_{\mu\nu}^{+}u_{\alpha} \right] D_{\beta}+{\rm H.c.} \right] N + d_9 \bar{N} \left[ \frac{1}{2m}i\epsilon^{\mu\nu\alpha\beta}\text{Tr} \left[  F_{\mu\nu}^{+}\right] u_{\alpha}D_{\beta}+{\rm H.c.} \right] N \nonumber \\
&+ d_{16} \bar{N} \left[ \frac{1}{2}\gamma^{\mu}\gamma_{5} \text{Tr} \left[ \chi_{+}\right] u_{\mu} \right] N + d_{18} \bar{N} \left[ \frac{1}{2}\,i\gamma^{\mu}\gamma_{5}[D_{\mu},\chi_{-}] \right] N \nonumber \\
&+ d_{20} \bar{N} \left[ -\frac{1}{8m^{2}} i\gamma^{\mu}\gamma_{5}[\widetilde{F}_{\mu\nu}^{+},u_{\lambda}]D^{\lambda\nu}+{\rm H.c.} \right] N  \nonumber \\
&+ d_{21} \bar{N} \left[ \frac{1}{2}i\gamma^{\mu}\gamma_{5} [\widetilde{F}_{\mu\nu}^{+},u^{\nu}] \right] N + d_{22} \bar{N} \left[ \frac{1}{2}\,\gamma^{\mu}\gamma_{5}[D^{\nu},F_{\mu\nu}^{-}] \right] N \,,
\end{align}
where $d_j$ ($j=8,9,16,18,20,21,22$) are new LECs appearing at $O(p^3)$ in units of ${\rm GeV}^{-2}$. The derivative operator $D^{\lambda\nu} = D^\lambda D^\nu + D^\nu D^\lambda $ acts over the  nucleon doublet~\footnote{The totally antisymmetric Levi-Civita tensor can be written as $\epsilon^{\mu\nu\alpha\beta}=-\frac{i}{8}\left[ \left\{ \left[ \gamma^\mu , \gamma^\nu \right], \gamma^\alpha \right\} ,\gamma^\beta \right] \gamma_5$.} and
\begin{equation}
 \widetilde{F}_{\mu\nu}^{+}=F_{\mu\nu}^{+}-\frac{1}{2}{\rm Tr}[F_{\mu\nu}^{+}].
\end{equation}

The interaction between the nucleon and $\Delta$ is described  by a Lagrangian that decouples the spin-$1/2$ components from the spin-$3/2$ Rarita Schwinger field~\cite{Pascalutsa:2007yg, Pascalutsa:2006up}. For a calculation up through $O(p^3)$ in the $\delta$ counting the relevant terms are
\begin{equation}
\mathcal{L}_{\Delta \pi N}^{(1)}=\frac{i h_A}{ 2 F m_\Delta} \bar{N} T^a \gamma^{\mu \nu \lambda}  (\partial_\mu \Delta_\nu) \partial_\lambda \pi^a + \text{H.c.},
\end{equation}
\begin{equation}
\mathcal{L}_{\Delta \gamma N}^{(2)}=\frac{3 i e g_M}{2m(m+m_\Delta)} \bar{N} T^3 (\partial_\mu \Delta_\nu) \tilde{f}^{\mu \nu} + \text{H.c.},
\end{equation}
where  $ \gamma^{\mu \nu \lambda}= \frac{1}{4} \left \{ \left[ \gamma^\mu , \gamma^\nu \right], \gamma^\lambda \right \}$, 
$\tilde{f}^{\mu \nu}=\frac{1}{2} \epsilon^{\mu \nu \alpha \beta} 
( \partial_\alpha A_\beta - \partial_\beta A_\alpha)$. Furthermore,  
 $\Delta_\nu =(\Delta^{++}_\nu, \Delta^+_\nu,\Delta^0_\nu,\Delta^-_\nu)^T $ are the components of the spin-$3/2$ Rarita Schwinger field corresponding to the isospin multiplet for the $\Delta$ resonance. The isospin transition matrices $T^a$ can be found in Ref.~\cite{Pascalutsa:2006up}. 
 
\subsection{Theoretical model}

\begin{figure}
\begin{center}
\includegraphics[width=0.9\textwidth]{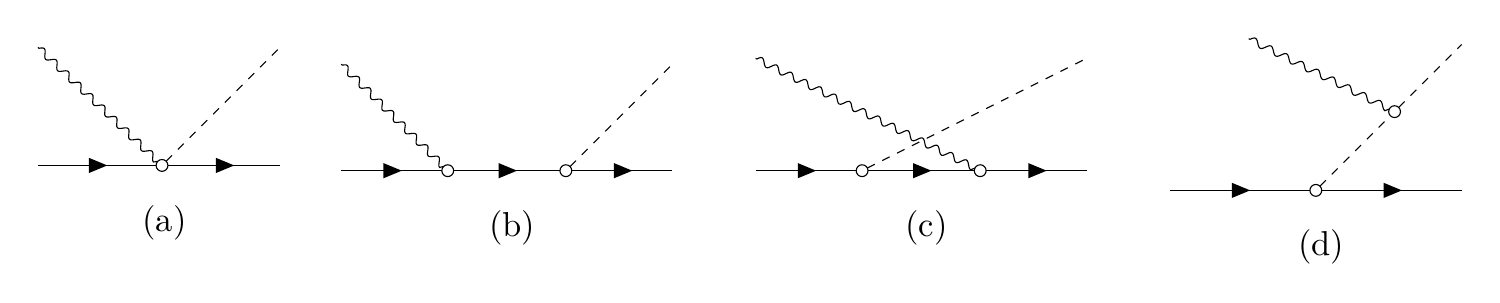}
\caption{Topologies of tree-level Feynman diagrams for the reaction $\gamma N \rightarrow \pi N'$. Diagrams (a)-(d) correspond to the $\mathcal{O}(p^k)$ amplitudes $\mathcal{T}_{(a)}^{(k)}$ -
$\mathcal{T}_{(d)}^{(k)}$ given in Appendix A. 
}
\label{fg:feynmanTree}
\end{center}
\end{figure}

\begin{figure}
\begin{center}
\includegraphics[width=0.6\textwidth]{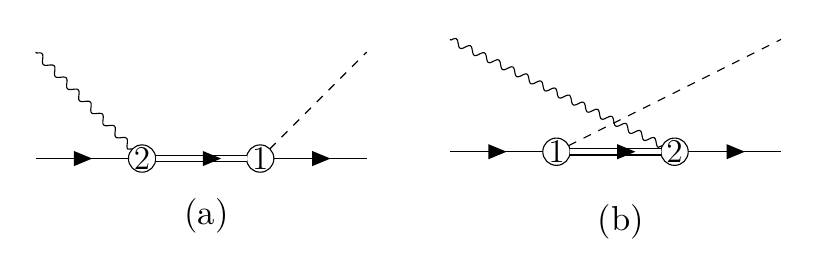}
\caption{Direct (a) and crossed (b) Feynman diagrams for the reaction $\gamma N \rightarrow \pi N'$ including the $\Delta$ resonance. The labels in the circles specify the chiral order for each vertex.}
\label{fg:feynmanDelta}
\end{center}
\end{figure}

The tree level Feynman diagrams contributing to the scattering amplitude up through order $O(p^3)$  are depicted in  Fig.~\ref{fg:feynmanTree} for the nucleonic sector, and in Fig.~\ref{fg:feynmanDelta} for the $\Delta(1232)$ resonance part.  The explicit expressions of the amplitudes are given in Appendix~\ref{appA}.

\begin{figure}
\begin{center}
\includegraphics[width=0.9\textwidth]{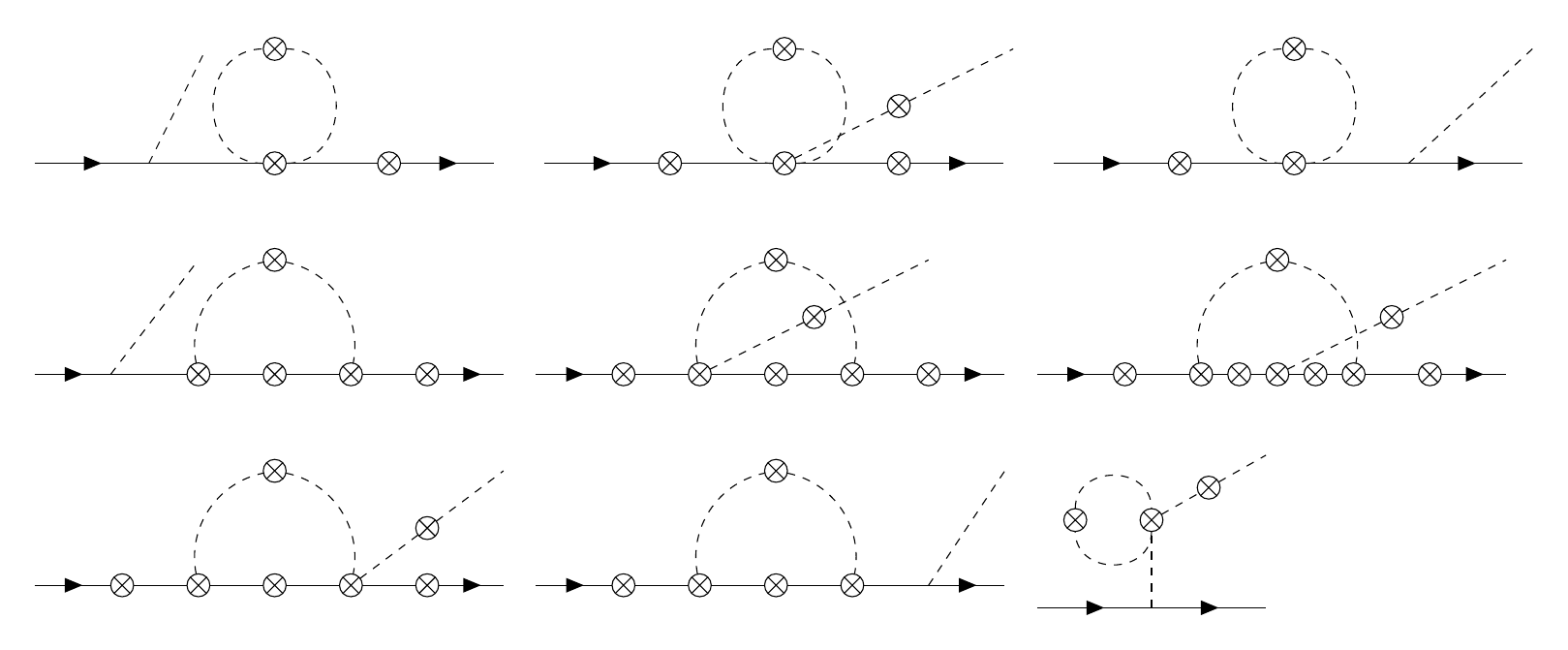}
\caption{One-loop topologies for the reaction $\gamma N \rightarrow \pi N'$ from which Feynman diagrams are generated. Solid lines indicate nucleons, dashed lines stand for pions and the crossed-circle vertices denote the position where incoming photons can be inserted. The topologies that lead to corrections in the external pion and nucleon legs are not shown because they are taken into account by  the wave function renormalization.}
\label{fg:feynmanLoop}
\end{center}
\end{figure}

Additionally to the tree diagrams,  we need the one loop amplitudes generated by the topologies shown in Fig.~\ref{fg:feynmanLoop}.
The calculation of the amplitudes has been carried out in {\it Mathematica} with the help of the FeynCalc package~\cite{Shtabovenko:2016sxi,Mertig:1990an}. The analytical results are very lengthy and not shown here but can be obtained from the authors upon request. \footnote{Expressions for the less general case of the $\gamma p\rightarrow p \pi^0$ process can be found in Ref.~\cite{Blin:2016itn}}

The ultraviolet (UV) divergences stemming from the loops are subtracted using the modified minimal subtraction scheme, i.e. $\widetilde{\rm MS}$ or equivalently $\overline{\rm MS}$-1,  
and here the renormalization scale $\mu$ is taken to be the nucleon mass. \footnote{In $\widetilde{\rm MS}$, one subtracts multiples of  $R=\gamma_E-1/\epsilon_{UV} -\log (4\pi)-1$, where $\epsilon_{UV}=(4-d)/2$ with $d$ the dimension of spacetime,
and $\gamma_E$ is the Euler constant.
}

To restore the power counting, we apply the EOMS scheme. Therefore, after the cancellation of the UV divergences we proceed to perform the required finite shifts to the corresponding LECs, so that the transformed parameter $\tilde X$ fulfills
\begin{align}
    X= \tilde X + \frac{ m \, \tilde\beta_X}{16 \pi^2 F^2}, 
\end{align}
which in our case applies for $X \in \{ m,g,c_1,c_6,c_7\}$.
For the parameters, $m$ and $g$, from $\mathcal{L}_{\pi N}^{(1)}$, we get
\begin{align}
 \tilde\beta_{m} =  - \frac{3}{2} g^2 \bar{A}_0\left[m^2\right], \qquad
 \tilde\beta_g = g^3 m + \frac{\left(2-g^2\right) g }{m} \bar{A}_0\left[m^2\right],
\end{align}
where
\begin{equation}
\bar{A}_0[m^2]    = -m^2 \log\frac{m^2}{\mu^2}
\end{equation}
is the ${\widetilde{\rm MS}}$-renormalized scalar 1-point Passarino-Veltman function with $\mu$ the renormalization scale introduced in the dimensional regularization.
For the second order LECs we 
have~\cite{Fuchs:2003ir}~\footnote{Note that  the EOMS shifts  applied to the $c_6$ and $c_7$ parameters in Ref.~\cite{Fuchs:2003ir} are different, since their Lagrangian has an alternative arrangement so that: $c_6=4m c_6^F, c_7=m\left(c_7^F - 2 c_6^F \right)$, where the superscript $F$ is just to identify the LECs in Ref.~\cite{Fuchs:2003ir}. } 
\begin{align}
\tilde \beta_{c_1} = \frac{3}{8} g^2 + \frac{3g^2}{8m^2} \bar{A}_0[m^2], \qquad
\tilde \beta_{c_6} = - 5 g^2 m, \qquad  
\tilde \beta_{c_7} = 4 g^2 m .
\end{align}

Finally, the full amplitude, $\mathcal{T}$, is related to the amputated one, $\hat{\mathcal T}$,  via the Lehmann-Symanzik-Zimmermann (LSZ) reduction formula \cite{Lehmann:1954rq}
\begin{align}
    \mathcal{T} = \mathcal{Z}^{\frac12}_\pi \mathcal{Z}_N \hat{\mathcal T},
\end{align}
where $\mathcal{Z}_\pi$ and $\mathcal{Z}_N$ are the wave function renormalization constants of the pion and nucleon, respectively. Their explicit expressions are given in Appendix \ref{AppB}.

\section{Fit procedure and error estimation}
\label{sec:fiterr}

\subsection{Experimental database}
We compare our theoretical model to the data in the energy range from threshold $\sqrt s\sim 1080$~MeV up to  $\sqrt s\sim 1130$~MeV. This choice guarantees that  the momentum of the outgoing pion is small and that we stay well below the $\Delta$ resonance peak. We should point out that we work in the isospin limit, both in the choice of the Lagrangian and in the further calculation of the loops. Therefore, the framework is not well suited for the measurements corresponding to the first MeV's above threshold, where the mass splittings are quite relevant.  We have checked, nonetheless, that our numerical results are not modified by the inclusion or exclusion of those data points.

%$\pi 0$
The larger part of the database corresponds to the $\gamma p\rightarrow \pi^0 p$ process. Furthermore, the experimental errors are relatively smaller   when compared to the other channels. As a consequence, the neutral pion production has a preeminent weight in the fits. There have been  extensive measurements in the near threshold region~\cite{Fuchs:1996ja,Bergstrom:1996fq,Bergstrom:1997jc,Schmidt:2001vg}, although the largest contribution comes from the comprehensive set of data on angular cross sections and photon asymmetries obtained at MAMI~\cite{Hornidge:2012ca}.~\footnote{The data from Refs.~\cite{Bergstrom:1996fq,Bergstrom:1997jc} are not unfolded from the angular spectrometer distortion and have not been included in the fit.} 
At the higher end of our energy range there are a few data points measured by the LEGS facility at the Brookhaven National Laboratory~\cite{Blanpied:2001ae}.

% 762 data points -80% of the total database

%****  pi-

In comparison, at these energies data are  scarce for the channels with charged pions and there are very few recent experiments on them. For the reaction $\gamma n\rightarrow \pi^- p$, we use the angular distributions and total cross sections from Refs.~\cite{Rossi:1973wf,Benz:1974tt,Salomon:1983xn,Bagheri:1987kf}. There are no data on polarization observables.
The early experiments at Frascati~\cite{Rossi:1973wf} and DESY~\cite{Benz:1974tt} actually measured the reaction on deuterium and then, the cross sections on the neutron were obtained using the spectator model.
On the other hand, the experiments at TRIUMF~\cite{Salomon:1983xn,Bagheri:1987kf} correspond to the inverse reaction: radiative pion capture on the proton. There are some later measurements  from the early 1990s, also at TRIUMF, quoted by SAID~\cite{SAID}, but they are unfortunately unpublished. Only recently, the $\pi^-$ photoproduction on the deuteron has been measured again at the MAX IV Laboratory~\cite{Strandberg:2018djk}, but the neutron cross section   has not been derived yet.

% 69 points total

There are some more data for the $\gamma p\rightarrow \pi^+ n$  channel, which can be measured more directly. They are  mostly angular and total cross sections but they also include some photon asymmetries. We take the data from Refs.~\cite{Walker:1963zzb,Fissum:1996fi,Blanpied:2001ae,Ahrens:2004pf}.

%126 points total 

In total, the database contains 957 points.
%\it{902 points if we consider photon energies above 150 MeV.} 
For most of them the total error estimation (statistic plus systematic) was given in the original references.
A typical 5\% systematic error has been added in quadrature for the few points where only the statistical error was provided \cite{Rossi:1973wf,Benz:1974tt,Salomon:1983xn}.

%%%%% 

\subsection{Low energy constants\label{sec:LECs}}
Most of the parameters required in the calculation are readily available as they have been obtained in the analysis of other processes or they are known functions of  physical quantities. The constants $g$, $F_0$, $m$  appearing in the lowest order terms of the Lagrangian
are given as a function of their corresponding physical values in Appendix~\ref{AppB}.
For the physical magnitudes we take $F = 92.42$~MeV, $g_A=1.27$, $m_\Delta=1232$~MeV and $e^2=4\pi/137$. 
\begin{table}[ht]
\caption{ Values of the LECs determined from other processes.}\label{tab:LECs}
\vspace{-0.5cm}
\begin{eqnarray}
\begin{array}{cc|rc}
\hline\hline
&{\rm LEC}& \text{Value}&\text{Source}\\
\hline
\multirow{2}{*}{$\mathcal{L}_{ N}^{(2)}$}&\tilde{c}_6&5.07\pm0.15&\text{$\mu_p$ and $\mu_n$~\cite{Bauer:2012pv,Yao:2018pzc,Patrignani:2016xqp}}\\
\cline{4-4}
&\tilde{c}_7&-2.68\pm0.08&\text{$\mu_p$ and $\mu_n$~\cite{Bauer:2012pv,Yao:2019avf,Patrignani:2016xqp}}\\
\hline
\multirow{2}{*}{$\mathcal{L}_{N}^{(3)}$}&d_{18}&-0.20\pm0.80 \; \text{GeV}^{-2}&\multirow{1}{*}{\text{$\pi N$~scattering~\cite{Alarcon:2012kn}}}\\
\cline{4-4}
&d_{22}&5.20\pm0.02 \; \text{GeV}^{-2}&\text{$\langle r_A^2\rangle_N$~\cite{Yao:2017fym}}\\
%d_{23}&\\
\hline
\mathcal{L}_{\pi N\Delta}^{(1)}&h_A&2.87\pm0.03&\text{$\Gamma_\Delta^{\rm strong}$~\cite{Bernard:2012hb}}\\
\hline
\mathcal{L}_{\gamma N\Delta}^{(2)}&g_M&3.16\pm0.16&\text{$\Gamma_\Delta^{\rm EM}$~\cite{Blin:2015era}}\\
\hline\hline
\end{array}\nonumber
\end{eqnarray}
\end{table}

In Table~\ref{tab:LECs}, we show the values of LECs obtained with the same framework (EOMS scheme + explicit $\Delta$) and at the same order [$O(p^3) $ in the $\delta$ counting] as the present work.~\footnote{In some of the  references, e.g.\ ~\cite{Bauer:2012pv}, the $\Delta$ resonance was not explicitly included, but its contribution starts at a higher order in the $\delta$ counting.} Apart from them, our theoretical model depends on the LECs $d_8$, $d_9$, $d_{16}$, $d_{20}$ and $d_{21}$. In our case $d_{16}$ can be absorbed by 
$g$ as shown in Appendix~\ref{appA}.~\footnote{The parameter $d_{16}$ has been investigated, within the current approach, studying the dependence on the pion mass of the axial coupling of the nucleon in lattice data. A value of $d_{16}=(-0.83\pm 0.03)$ GeV$^{-2}$ was obtained in Ref.~\cite{Yao:2017fym}.} The remaining four constants have been fitted to the experimental data minimizing the $\chi$ squared.

\subsection{Error estimation}
There are two sources of uncertainties in our prediction of any observable. First,  there is  the  uncertainty propagated from the statistical errors of the LECs in the fit, which we take as
\begin{align}
\label{eq:corr}
    \delta \mathcal{O}_{\rm LECs}=\left( \sum_{i,j}  \left[ \text{Corr}(x_i,x_j) \right] 
     \frac{\partial \mathcal{O}(\bar{x}_i)}{\partial x_i} \delta x_i
      \frac{\partial \mathcal{O}(\bar{x}_j)}{\partial x_j} \delta x_j
    \right)^{1/2},
\end{align}
where 
$\mathcal{O}$ refers to the observable, and the $i$ and $j$ indices are labels for a given LEC $x_i$,  with $\bar{x}_i$ and $\delta{x}_i$ its corresponding mean and error values as obtained from the fit. Finally, $\text{Corr}(i,j)$ is the $(i,j)$th matrix element of the correlation matrix.  

Additionally, we consider the systematic errors due to the truncation of the chiral series.
 We have used the method from Refs.~\cite{Epelbaum:2014efa,Siemens:2016hdi} where the uncertainty
 $\delta \mathcal{O}^{(n)}_{th}$, at order $n$ for any observable $\mathcal{O}$  is given by
\begin{align}
    \delta \mathcal{O}^{(n)}_{th}=\text{max} \left( \left| \mathcal{O}^{(n_{LO})} \right| Q^{n-n_{LO}+1}, \left\{ \left| \mathcal{O}^{(k)} - \mathcal{O}^{(j)}\right| Q^{n-j} \right\} \right), \qquad n_{LO} \leq j \le k \leq n\,,\label{Eq:order}
\end{align}
 where $Q=m_\pi/\Lambda_{b}$, $\Lambda_b$ is the breakdown scale of the chiral expansion. We set $\Lambda_b=4\pi F \sim 1$ GeV as in Ref.~\cite{Yao:2017fym}. In our case, the lowest order considered is $n_{LO}=1$ and the upper order calculated is $n=3$.

\section{Results and discussion}
\label{sec:res}
\subsection{Fit with and without $\Delta$ contribution}
We  have fitted the free LECs of our model comparing our calculation with the experimental database and minimizing the $\chi$ squared. The results of the fit for several different options are given in Table~\ref{tab:LECfit}. 
Fit I corresponds to our full model, as described in the previous sections. The LECs from Table~\ref{tab:LECs} have been set to their central values except for $d_{18}$, which has been left to vary within the quoted range, and $g_M$ that was left free. 
In the minimization procedure, we have chosen the combinations $d_8+d_9$ and $d_8-d_9$ because of the strong correlation existing between  $d_8$ and $d_9$
which cannot be well determined independently. Furthermore, the channel $\gamma p\rightarrow \pi^0 p$, with the most accurate data, depends just on $d_8+d_9$, which leads to a quite precise value for this combination. With the current data, we are less sensitive to $d_8-d_9$ that would benefit from better data on the other channels that depend only on $d_9$ for the $\gamma p\rightarrow\pi^+ n$ and $\gamma n\rightarrow\pi^- p$ cases, and on $d_8-d_9$ for $\gamma n\rightarrow\pi^0 n$.

Also, the $O(p^3)$ constants  $d_{20}$ and $d_{21}$ are less constrained, because they only affect the channels with charged pions. These latter channels are already relatively well described by lower order calculations and are not very sensitive to third order effects. Furthermore, the uncertainties in their data are comparatively larger than for the $\pi^0$ channel. 
We must also recall that, in pion photoproduction, $d_{21}$ is fully correlated with $d_{22}$ and only appears in the amplitudes in the $2d_{21}-d_{22}$ combination. Thus, the value shown in Table~\ref{tab:LECfit} depends straightforwardly on $d_{22}$.

\begin{table*}[t]
\caption{
The values of the LECs are dimensionless for 
 $g_M$ and in units of GeV$^{-2}$ for $d$'s.  Fit I refers to the standard setting, Fit II removes $\Delta$ mechanisms, Fit III leaves $d_{18}$ free. In Fits I and II,   $d_{18}$ is restricted to the 1 $\sigma$ range given in Table~\ref{tab:LECs}, and therefore shown in boldface.  }\label{tab:LECfit}
\begin{tabular}{l | c c c}
LECs & Fit I & Fit II - $\slashed \Delta$ & Fit III \\
\hline
$d_8 + d_9$ & $1.16 \pm 0.01$& $ 3.53 \pm 0.01$ & $ 0.95 \pm 0.02$ \\
$d_8 - d_9$ & $1.09 \pm 0.18$ & $5.31 \pm 0.24$& $0.27 \pm 0.18$ \\
$d_{18}$ & $\mathbf{0.60} $ &$\mathbf{-1.00} $ & $5.69 \pm 0.14$ \\
$d_{20}$ & $-0.74 \pm 0.17$ & $-3.81 \pm 0.19$  & $1.87 \pm 0.19$\\
$d_{21}$ & $4.32 \pm 0.14$ &$6.98 \pm 0.15$ &$4.58 \pm 0.15$ \\
$g_{M}$ & $2.90 \pm 0.01$ & - & $3.16 \pm 0.02$ \\
\hline
$\chi^2_{TOT}/dof$ & $\mathbf{3.22}$ & $\mathbf{29.5}$ & $\mathbf{1.58}$ \\
\hline
$\chi^2_{\pi 0}/dof $  & $3.58$  & $37.2$ & $1.31$ \\
$\chi^2_{\pi +}/dof$ & $1.89$ & $1.79$ & $2.43$ \\
$\chi^2_{\pi -}/dof$ & $1.99$ & $1.90$ & $2.76$ \\
\hline
\end{tabular}
\end{table*}

\begin{figure} [ht]
\begin{center}
\includegraphics[width=0.65\textwidth]{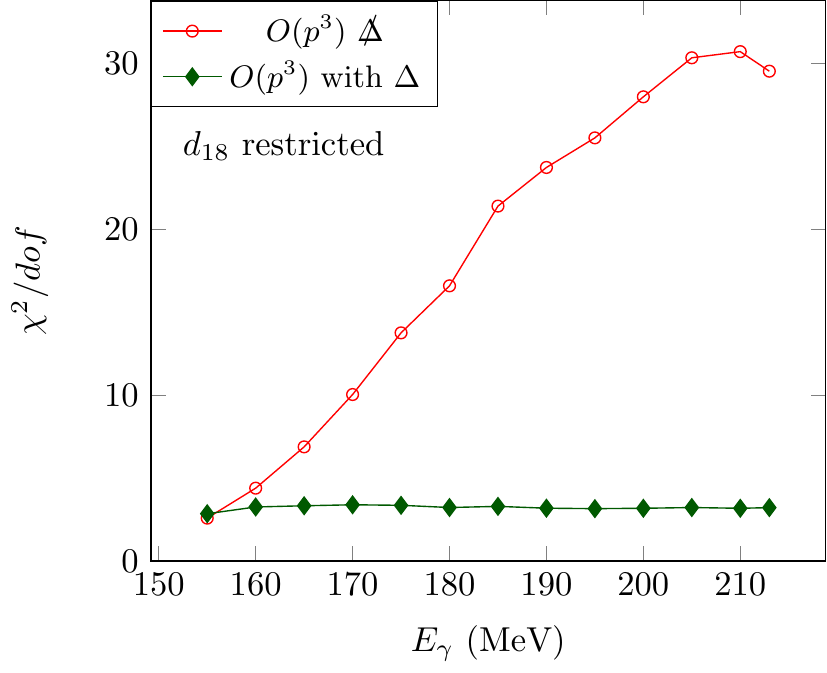}
\caption{$\chi^2$ per degree of freedom as a function of the maximum photon energy of the data included in the fit. Solid diamonds:
full model at order $O(p^3)$ with $\Delta$ resonance, empty circles: model without the $\Delta$ inclusion. Lines to guide the eye.}
\label{fg:chi2}
\end{center}
\end{figure}

A first remark is that $g_M$ takes a value consistent with that obtained from the electromagnetic $\Delta$ decay width. This clearly shows the sensitivity of the pion photoproduction to the $\Delta$ resonance even at the low energies investigated. In fact, removing the $\Delta$ mechanisms we get Fit II, with a much worse agreement with data.
 The reshuffling of  the free parameters is ineffective in describing the rapid growth of the cross section of the $\pi^0$ channel. The importance of the resonant mechanisms can be also appreciated in Fig.~\ref{fg:chi2}. The quality of the agreement decreases rapidly as a function of the maximum photon energy of the data included in the fit in the $\Delta$-less case, whereas it is practically stable for the full model. This behavior (rapid growth of $\chi^2$ as a function of energy) can also be seen even for $O(p^4)$ covariant and HB calculations that do not include the $\Delta$ resonance explicitly. See, e.g., Figs. 2 of Refs.~\cite{Hilt:2013uf,Hilt:2013fda} and Fig.~1 of Ref.~\cite{FernandezRamirez:2012nw}~\footnote{The two figures from Refs.~\cite{FernandezRamirez:2012nw,Hilt:2013uf} only consider the $\pi^0$ channel, whereas Fig.~\ref{fg:chi2} includes all the channels. Still, the comparison is fair as the $\chi^2$ is basically driven by the $\pi^0$ channel and we obtain a similar figure for that restricted case.}. 

Comparing the absolute values of $\chi^2/dof$ ($\chi$ squared per degree of freedom), we see that the $O(p^3)$ calculation without $\Delta$ (Fit II) gives $\chi^2/dof=29.5$. This number is mostly driven by the contribution of the $\gamma p \rightarrow \pi^0 p$ channel (Table~\ref{tab:LECfit}),   whereas the contribution of the channels with charged pions to $\chi$ squared is barely modified. The $\chi^2$  value is substantially reduced with the explicit inclusion of the $\Delta$ (Fit I), still at $O(p^3)$ and even when the corresponding LECs are previously fixed. A  reduction can also be obtained without the $\Delta$ by doing an $O(p^4)$ calculation~\cite{FernandezRamirez:2012nw,Hilt:2013uf}. However, apart from requiring a number of extra parameters, in the $\Delta$-less calculations the fit quality diminishes rapidly as a function of the photon energy.

\begin{figure}[ht]
\begin{center}
\includegraphics[width=1\textwidth]{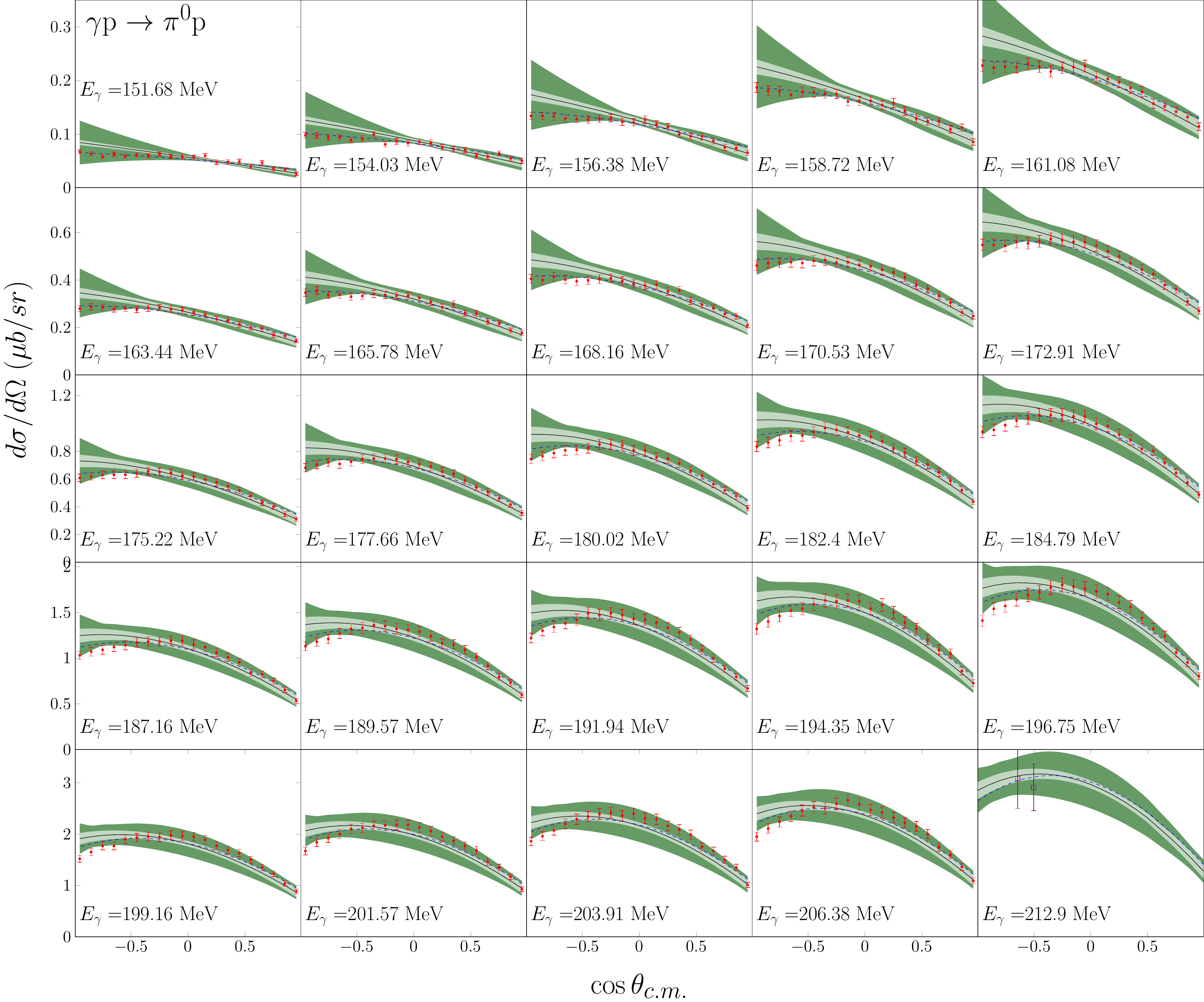}
\caption{Angular cross section for the $\gamma p \rightarrow \pi^0 p$ channel at various energies. Solid line: theoretical model (Fit I).  Dashed line: Fit III. Data from Ref.~\cite{Hornidge:2012ca} marked as red points and from \cite{Blanpied:2001ae} as violet squares.  The inner band represents the statistical errors obtained by varying the LECs within their  uncertainties (as shown in Table~\ref{tab:LECs}) in the fit. The outer band stands for the total errors where the theoretical uncertainties (due to the chiral truncation) are added to the statistical ones in quadrature. 
}
\label{fg:dsgpi0p}
\end{center}
\end{figure}

\begin{figure}[ht]
\begin{center}
\includegraphics[width=1\textwidth]{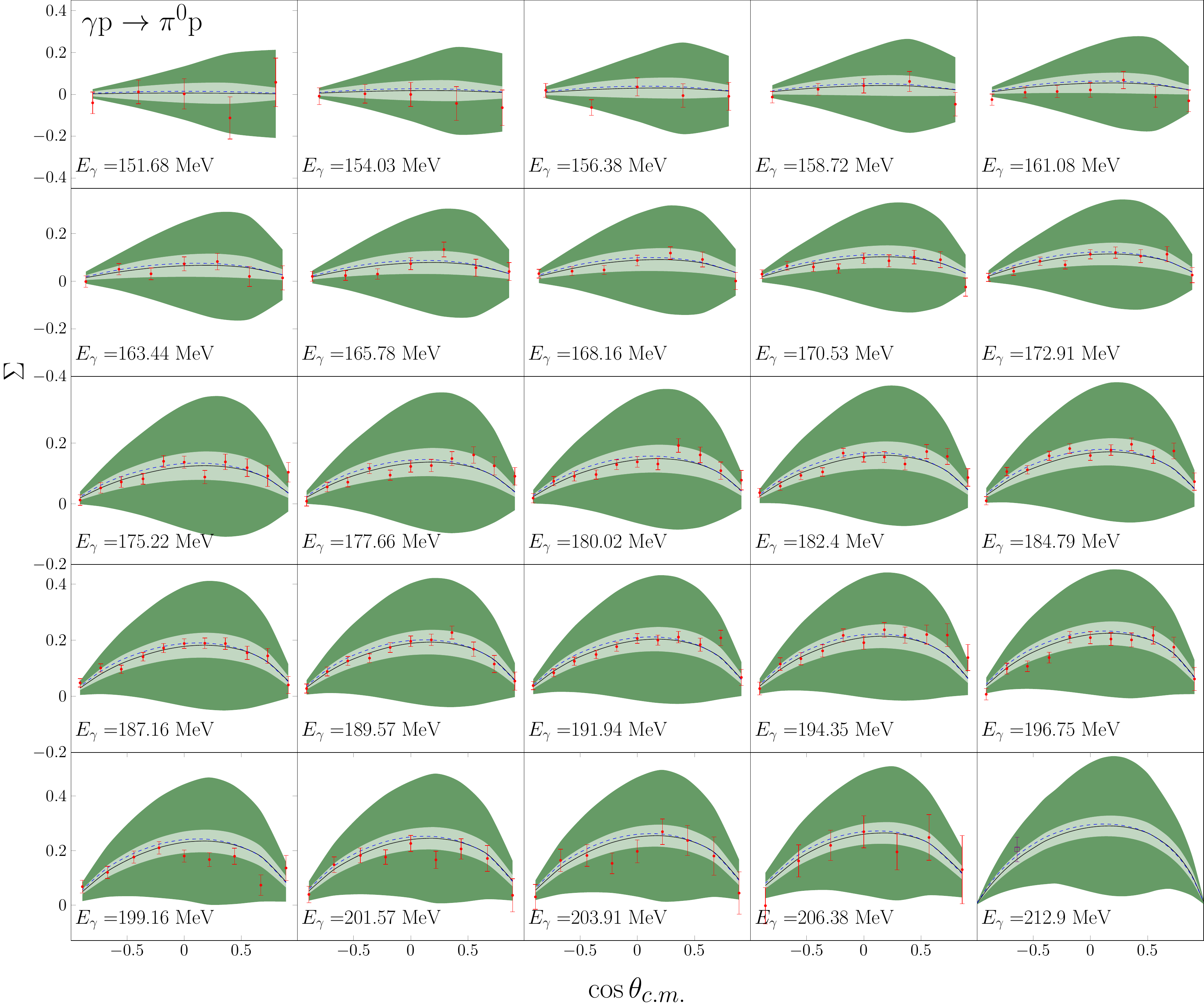}
\caption{Beam asymmetry for the $\gamma p \rightarrow \pi^0 p$ channel at various energies. Data from Ref.~\cite{Hornidge:2012ca} marked as red points and from \cite{Blanpied:2001ae} as a violet square.
Description same as Fig.~\ref{fg:dsgpi0p}.
}
\label{fg:asympi0p}
\end{center}
\end{figure}

\begin{figure} [ht]
\begin{center}
\includegraphics[width=0.65\textwidth]{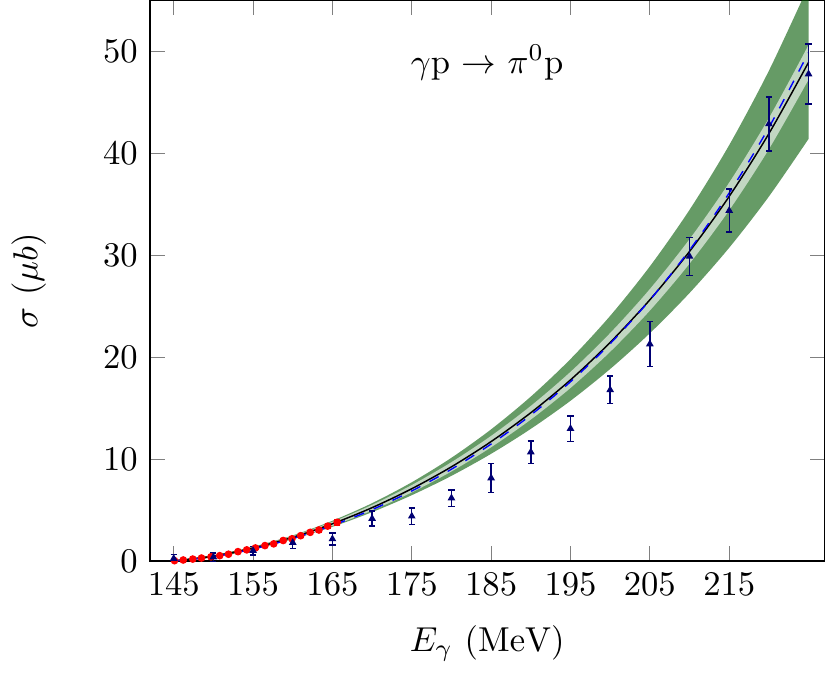}
\caption{Cross section for the $\gamma p \rightarrow \pi^0 p$ channel. Solid line: theoretical model, circles: data from Ref.~\cite{Schmidt:2001vg}, triangles: data from Ref.~\cite{Schumann:2010js}, not included in the fit. Description same as Fig.~\ref{fg:dsgpi0p}.}
\label{fg:sgtpi0p}
\end{center}
\end{figure}

In Figs.~\ref{fg:dsgpi0p}, \ref{fg:asympi0p} and \ref{fg:sgtpi0p}, we compare the results from Fit I with data from the $\pi^0$ channel. The only free third order LEC is the $d_8+d_9$ combination.
The agreement is overall good for both cross section and beam asymmetries in the full range of energies considered. Only the total cross sections from Ref.~\cite{Schumann:2010js} are systematically below the calculation from 165 to 205~MeV, see Fig.~\ref{fg:sgtpi0p}. However, these data are incompatible with the differential cross sections measured at the same energies in Ref.~\cite{Hornidge:2012ca}. Also, there is some overestimation (within the error bands but systematic) of the angular distributions at backward angles. The uncertainties due to the truncation of the chiral expansion are considerable. This fact reflects the large size of the $\Delta$ contribution and the $O(p^3)$ mechanisms to this observable.

\begin{figure}
\begin{center}
\includegraphics[width=0.8\textwidth]{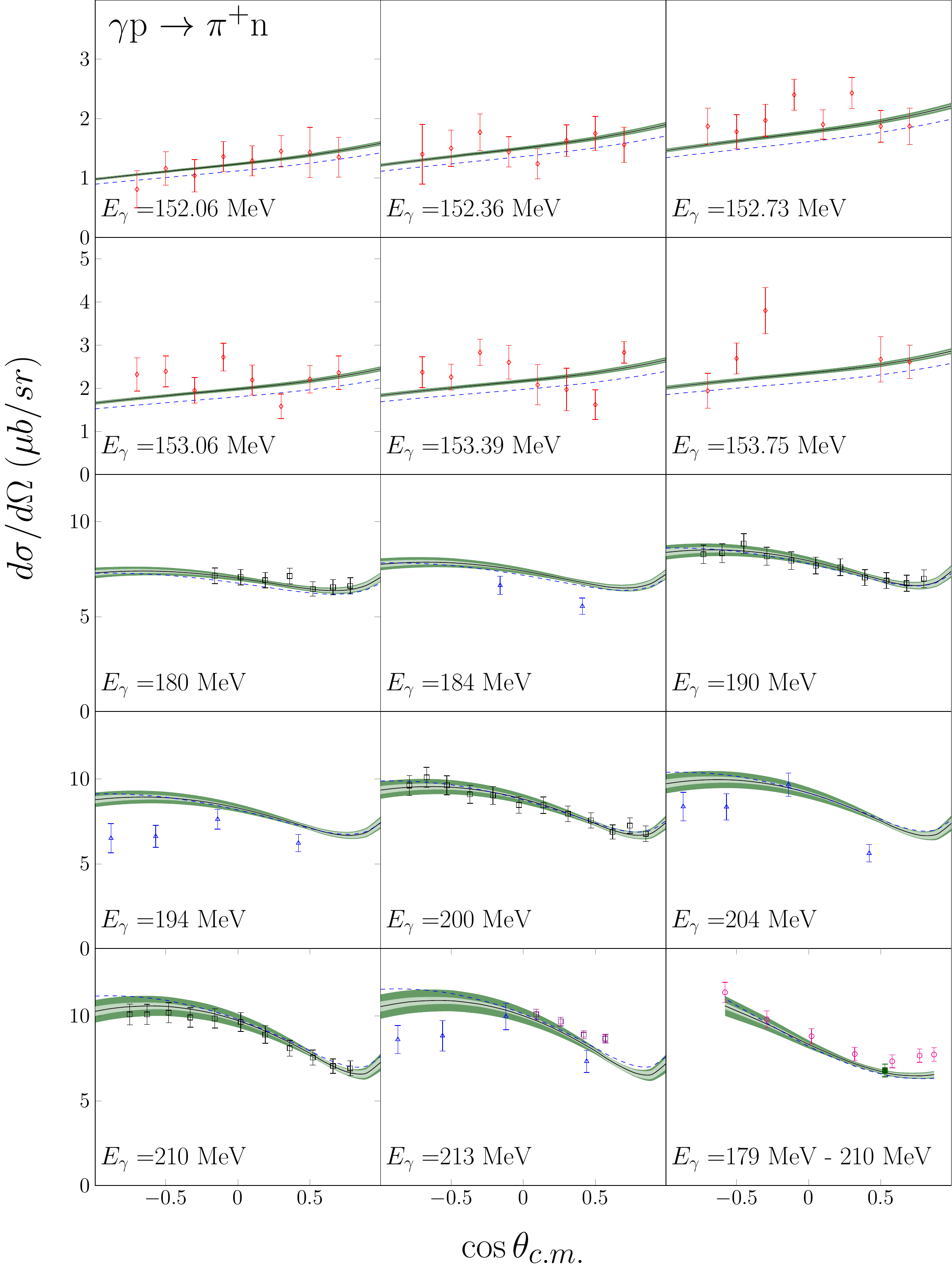}
\caption{Angular cross section for the $\gamma p \rightarrow \pi^+ n$ channel at various energies. Going from low to high energy, the data from Ref.~\cite{Korkmaz:1999sg} are marked as red diamonds, \cite{Ahrens:2004pf} as black squares, \cite{Fissum:1996fi} as blue triangles, \cite{Blanpied:2001ae} as violet squares. Finally, in the lowest right panel, data for different energies from Ref.~\cite{Walker:1963zzb} are marked as magenta circles and from \cite{Bergstrom:1996fq} as dark-green squares. In this latter panel the theory has been calculated at precisely the energies and angles of the data points, and the lines and bands have been interpolated.
Description same as Fig.~\ref{fg:sgtpi0p}.
}
\label{fg:dsgpipn}
\end{center}
\end{figure}
\begin{figure}
\begin{center}
\includegraphics[width=0.5\textwidth]{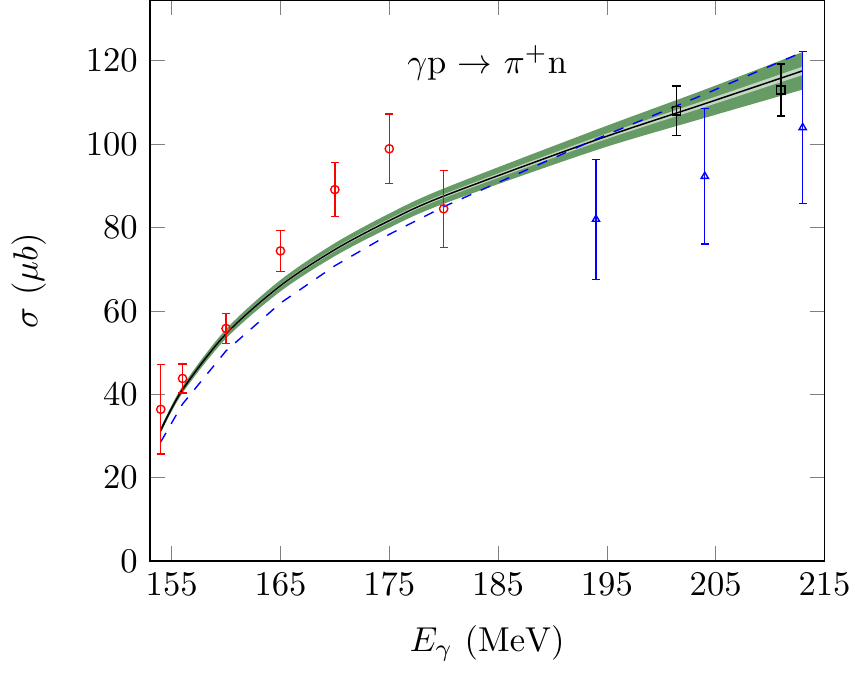}
\caption{Cross section for the $\gamma p \rightarrow \pi^+ n$ channel at various energies. Data from Ref.~\cite{McPherson:1964zz} presented as red circles. In the same way as in Fig.~\ref{fg:dsgpipn}, data from \cite{Fissum:1996fi} as blue triangles and \cite{Ahrens:2004pf} as black squares.
Description same as Fig.~\ref{fg:sgtpi0p}.
}
\label{fg:sgtpipn}
\end{center}
\end{figure}

\begin{figure}
\begin{center}
\includegraphics[width=0.5\textwidth]{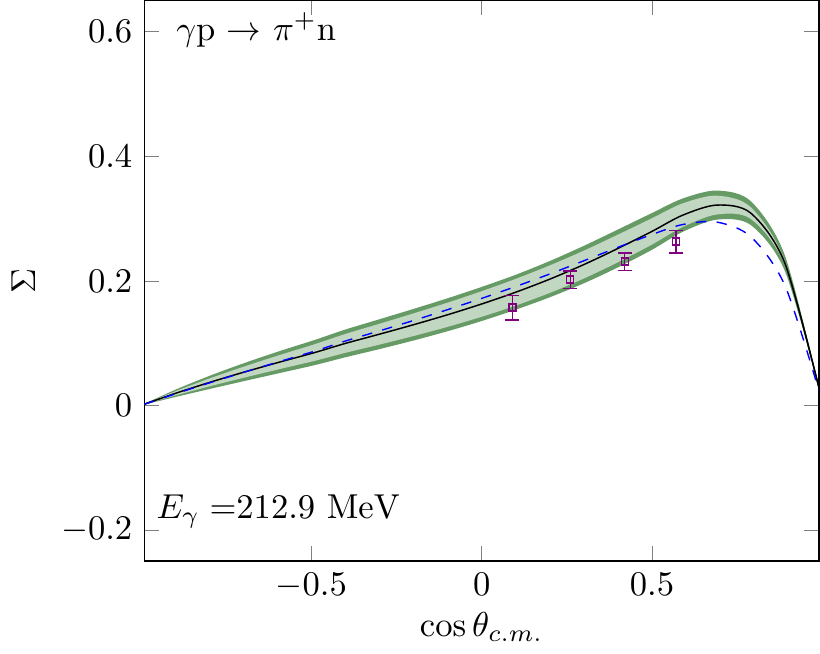}
\caption{Beam asymmetry for the $\gamma p \rightarrow \pi^+ n$ channel at $E_\gamma=212.9$ MeV. Data from Ref.~\cite{Blanpied:2001ae}.
Description same as Fig.~\ref{fg:sgtpi0p}.
}
\label{fg:asympipn}
\end{center}
\end{figure}
\begin{figure}
\begin{center}
\includegraphics[width=0.5\textwidth]{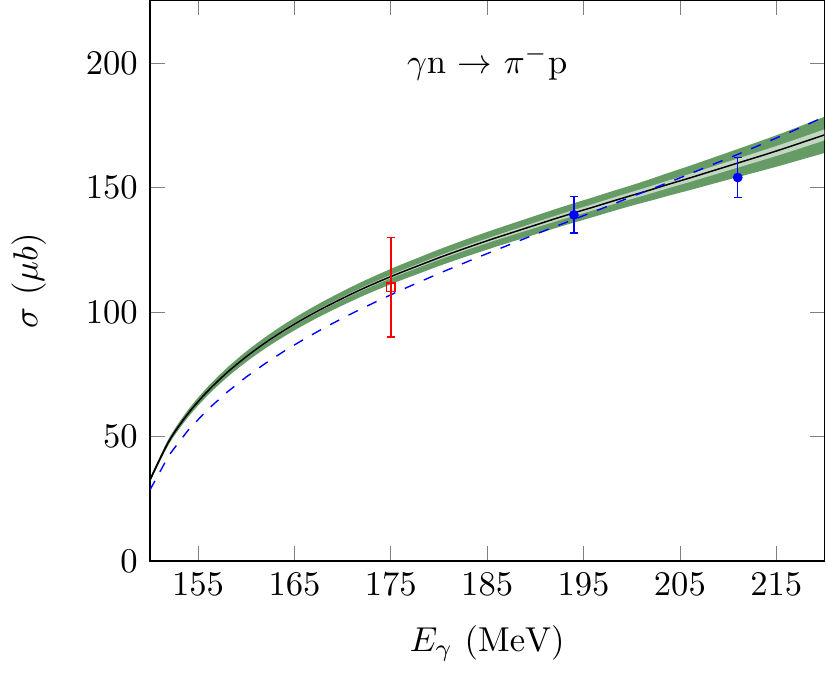}
\caption{Cross section for the $\gamma n \rightarrow \pi^- p$ channel at various energies. Data from Ref.~\cite{White:1960ukk} presented as red square and data from \cite{WA92} as blue filled circles.
Description same as Fig.~\ref{fg:sgtpi0p}.
}
\label{fg:sgtpimp}
\end{center}
\end{figure}
\begin{figure}
\begin{center}
\includegraphics[width=0.85\textwidth]{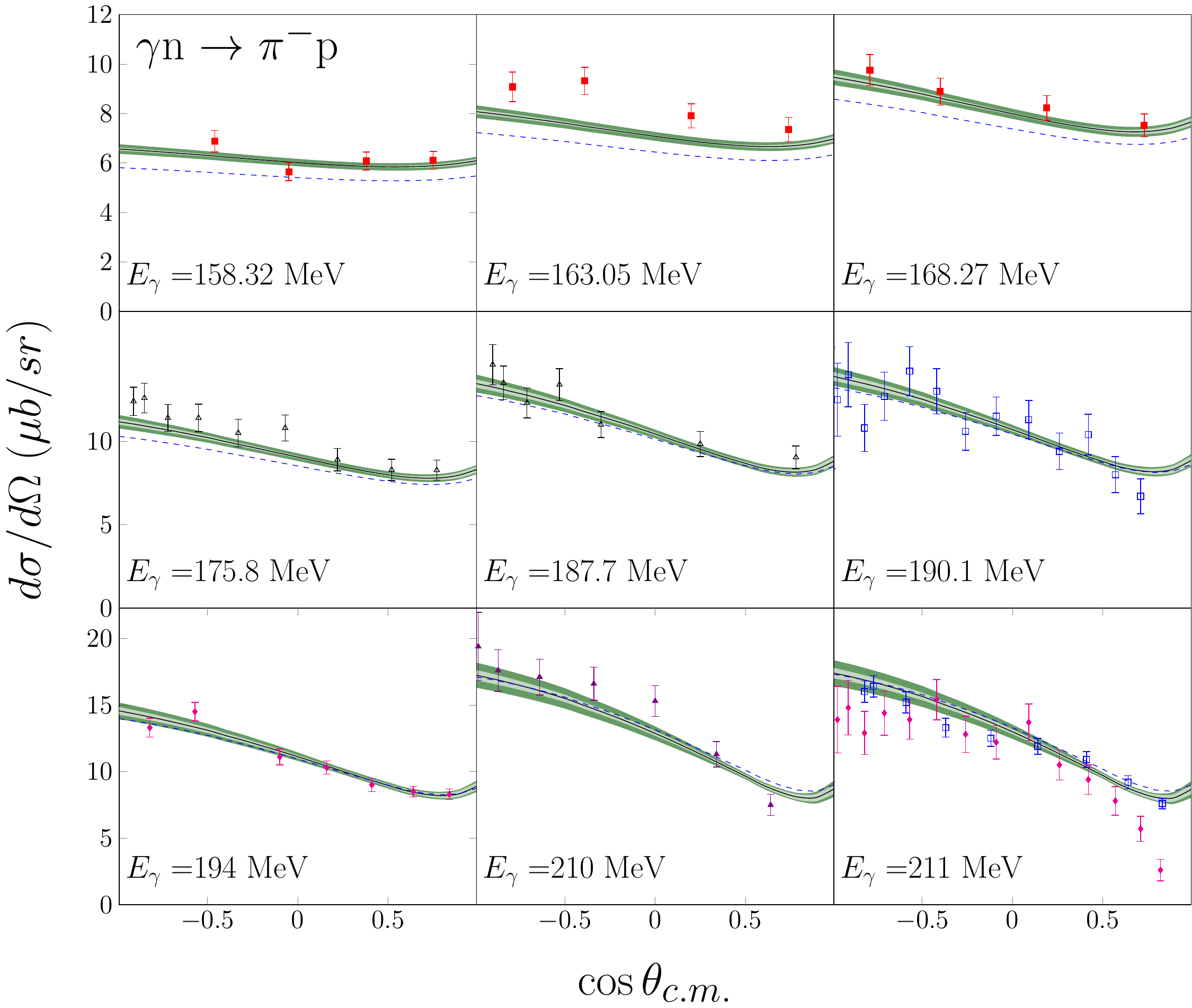}
\caption{Angular cross section for the $\gamma n \rightarrow \pi^- p$ channel at various energies. Data from Ref.~\cite{LI94} are presented as red filled squares, \cite{Salomon:1983xn} marked as black triangles, \cite{Rossi:1973wf} as blue squares, \cite{Benz:1974tt} as violet filled triangles and data from \cite{Bagheri:1987kf} as magenta filled diamonds.
Description same as Fig.~\ref{fg:sgtpi0p}.
}
\label{fg:dsgpimp}
\end{center}
\end{figure}

 The channel $\gamma p\rightarrow \pi^+ n$ is sensitive to the LECs $d_9$, $d_{20}$ and $d_{21}$. As shown in Figs.~\ref{fg:dsgpipn}, \ref{fg:sgtpipn} and \ref{fg:asympipn}, the agreement is good for the cross sections and for the few data available on beam asymmetry. The model also agrees well with the $\gamma n\rightarrow \pi^- p$ data
as shown in Figs.~\ref{fg:sgtpimp} and \ref{fg:dsgpimp}. This channel depends on the same third order LECs as the previous one.  The measurements in this channel are scarce and the uncertainties are relatively large. However, it gets a larger $\chi^2$ than the $\pi^+$ channel. This may come from some underestimation of the experimental  uncertainties. Actually, most of the contribution of this channel to the $\chi^2$ comes from regions with conflicting and incompatible measurements, such as the angular distribution at forward angles at $E_\gamma=211$~MeV.


The combination $d_8+d_9$ is very precisely determined in our fits as compared to the other third order LECs and, in particular, to $d_8-d_9$. Using the correlation matrix and Eq.~(\ref{eq:corr}), we can estimate their individual uncertainties. We obtain $d_8=1.13\pm 0.09$ GeV$^{-2}$ and $d_9=0.04\pm 0.09$ GeV$^{-2}$ for Fit I. The same values and uncertainties are obtained using directly $d_8$ and $d_9$, instead of their combinations, in the fit. These separate uncertainties are 1 order of magnitude larger than for the $d_8+d_9$ combination.

%\subsection{Higher chiral orders?}
We have explored the stability of the minimum of our standard result (Fit I) by removing the constraints previously imposed on $d_{18}$,  $c_6$ and $c_7$. Notice that $g_M$ was already free, $h_A$  always appears in a combination multiplied by $g_M$  and  $d_{22}$ is fully correlated with $d_{21}$. The results are shown in the Fit III of Table~\ref{tab:LECfit}. The $\chi^2$ improves substantially, mostly due to a better agreement with the $\pi^0$ channel and, in particular, the cross section at backward angles, Fig.~\ref{fg:dsgpi0p}. Notice, however, the slight worsening of the agreement with the charged pion channels.  While this might be pointing to some issue with the neutral pion production data at backward angles~\cite{Hornidge:2012ca}, it could also be arising from the theory side, as detailed below.

Of the now unconstrained LECs, the  $c_6$ and $c_7$ values remain in the range given in Table~\ref{tab:LECs}, but $d_{18}$  prefers positive values which are not acceptable as they are hardly compatible with the pion nucleon coupling constant $g_{\pi N}$~\cite{Alarcon:2012kn}.  
We have also examined how $\chi^2$ changes when moving $d_{18}$ across the  range $[-1.0,0.60]$  GeV$^{-2}$ given in Table~\ref{tab:LECs}. At  $d_{18}=-1.0$  GeV$^{-2}$, it  goes up to $\chi^2=4.70$.~\footnote{In Fit II, it rises up to $\chi^2=31.7$ at  $d_{18}=0.6$ GeV$^{-2}$.}
 
We have  found that the value $d_{18}$ changes much by modifications such as whether the wave function renormalization is applied to the full amplitude or the first order only, and whether the physical mass or $m_2$ [see Eq.~\eqref{eq:m2}] is used in the loops for the nucleons. All these options amount to $O(p^4)$ variations, and the fact that the value of $d_{18}$ is strongly affected by them may indicate the need for a higher order calculation to reach a proper chiral convergence.

A first step would be the inclusion of $O(p^{7/2})$ mechanisms, which correspond to tree mechanisms with  higher order $\Delta \pi$  or $\Delta \gamma$ coupling and a set of loop diagrams with $\Delta$ propagators inside the loop. This approach was already explored in Ref.~\cite{Blin:2016itn} for the $\pi^0$ channel and did  not change much the results as compared with the third order calculation, remaining consistent with the preference of large positive $d_{18}$ values. 

A full $O(p^4)$ calculation would incorporate further extra terms. The fourth order Lagrangian, $\mathcal{L}_N^{(4)}$, that contributes to the process entails fifteen additional couplings~\cite{Hilt:2013fda}.~\footnote{With the current dataset, the use of the full $\mathcal{L}_N^{(4)}$ Lagrangian with fifteen extra parameters and an already small $\chi^2$ leads to many minima and obvious overfitting.}
We have estimated the importance of this order by considering the tree level amplitude generated by $\mathcal{L}_N^{(4)}$. The explicit expression can be found in Appendix C from Ref.~\cite{Hilt:2013fda}. In particular, we have explored how $d_{18}$ is affected by the new terms and we have found that  it is very sensitive to some of the parameters,
as  $e_{48}$,  $e_{50}$ or   $e_{112}$.

\subsection{Convergence of the approach}
\begin{table*}[ht]
\caption{LECs and $\chi^2$ for calculations at different chiral orders. Bold numbers are fixed and depend only on physical quantities such as $g_A$, the proton and neutron magnetic moment and the $\Delta$ decay width. Thus, they  are not fitted to the pion photoproduction process.}
\label{tab:conv}
\begin{tabular}{l | c c c c}
LECs & $O(p^1)$ & $O(p^2)$ & $O(p^{5/2})$ & $O(p^3)$, Fit I \\
\hline
$g$ & $\bf 1.27$ & $\bf 1.27$ & $\bf 1.27$ & $\bf 1.11$
\\
$c_6$ & - & $\bf 3.706 $ & $\bf 3.706 $ & $\bf 5.07 $
\\
$c_7$ & - & $\bf -1.913 $ & $\bf -1.913 $ & $\bf -2.68 $
\\
$d_{18}$ & - & - & - & $\bf 0.60$ 
\\
$d_{22}$ & - & - & - & $\bf 5.20$
\\
$d_8 + d_9$ & - & - & - & $ 1.16 \pm 0.01$
\\
$d_8 - d_9$ & - & - & - & $1.09 \pm 0.18 $
\\
$d_{20}$ & - & - & - & $-0.74 \pm 0.17$
\\
$d_{21}$ & - & - & - & $4.32 \pm 0.14$
\\
$h_A$ & - & - & $\bf 2.87$ & $\bf 2.87$
\\
$g_{M}$ & - & - & $ \bf 3.16 $ & $2.90\pm 0.01$
\\
\hline
$\chi^2_{TOT}/dof$ & 165 & 310 & 60.7 & 3.22\\
\hline
$\chi^2_{\pi 0}/dof $  & 208 & 392 & 76.6 & 3.58\\
$\chi^2_{\pi +}/dof$ & 10.7 & 9.15 & 2.88 & 1.89\\
$\chi^2_{\pi -}/dof$ & 5.73 & 6.29 & 2.33 & 1.99\\
\hline
\end{tabular}
\end{table*}
In Table~\ref{tab:conv}, we show the $\chi^2$ results for the calculations at different chiral orders. At the lowest order, there is no free LEC. The amplitude only depends on physical magnitudes such as $g_A$, the masses, charges and the pion decay constant. The agreement is acceptable for the pion charged channels but quite bad for the $\pi^0$ one. The reason is well known as being due to the large cancellation between the different pieces of the $O(p)$ amplitude which leads to small cross sections and a large sensitivity to higher orders. The situation does not improve in a second order calculation. Again there are no free parameters. The new tree diagrams correspond to $c_6$ and $c_7$ terms which are directly connected to the magnetic moments of the neutron and proton.~\footnote{Actually, the cross sections are slightly better described, but there are strong disagreements with the beam asymmetry of the $\pi^0$ channel. }
Next, in the $\delta$ counting, comes the inclusion of the $\Delta$ mechanisms which start contributing at $O(p^{5/2})$. Once more, there are no free constants. We already get a much better description in the three channels. Still, the agreement is poor for the neutral pion channel.  An even larger improvement is reached in a third order calculation, without $\Delta$ but with some extra free parameters (Fit II of Table~\ref{tab:LECfit}). At this order, the loop diagrams start appearing. They are also important for improving the agreement of all channels. Finally, in the last column we show our Fit I results, incorporating both $\Delta$ mechanisms and a full third order calculation. It leads to an overall good agreement with the data in all channels. 

Altogether, the $\Delta$ mechanisms and the third order contributions play a capital role in reaching a good description of the pion photoproduction process. This is especially the case for neutral pion photoproduction, but also the charged pion production channels feel the improvement. We remark here that these effects also play a significant role  in weak pion production~\cite{Yao:2018pzc,Yao:2019avf}. 

\section{Summary\label{sec:summary}}

In this work, we have  investigated  pion photoproduction on the nucleon close to threshold in covariant ChPT, following the EOMS renormalization scheme.  Our approach includes explicitly the $\Delta(1232)$ resonance mechanisms. We have made a full calculation up through $O(p^3)$ in the $\delta$ counting.

The model reproduces well the total cross section, angular distributions and polarization observables for all the channels. The agreement is better, and for a wider range of energies, than in the $O(p^4)$ calculations in  both, covariant~\cite{Hilt:2013uf} and HB~\cite{FernandezRamirez:2012nw} schemes, without explicit $\Delta$. As in their case, our model without $\Delta$ only reproduces the data very close to threshold. This shows that the $\Delta$ resonance is instrumental in reproducing the energy dependence of the various observables. We should remark here that the $\Delta$ couplings are strongly constrained from its strong and electromagnetic widths.

With the simultaneous incorporation of all pion photoproduction channels,  our fit constrains some 
unknown  $O(p^3)$ LECs. Of these,  the combination $d_8+d_9$ is the most precisely determined due to the high quality of the $\gamma p\rightarrow p\pi^0$ data. 
The  constants  $d_{20}$, $d_{21}$ or $d_{9}$ separated from $d_8$, which only appear in the other channels involving charged pions, are not so well determined because data are scarce and typically with large uncertainties. New measurements on the $\gamma p\rightarrow n\pi^-$, $\gamma n\rightarrow p\pi^-$ or the reverse $n\pi^-\rightarrow\gamma p$ processes would be  useful to better pin down the values of these LECs. 
Finally, the extension to the description of electro- and weak production data will advance these studies even further, while offering the possibility of making reliable and accurate predictions for weak processes where data are more scarce.

\section*{Acknowledgments}
This research is supported by MINECO (Spain) and the ERDF (European Commission) grant No.  FIS2017-84038-C2-2-P, and SEV-2014-0398. It is also supported in part by the National Science Foundation of China under Grant No. 11905258, by the Fundamental Research Funds for the Central Universities under Grant No. 531118010379 and by the Deutsche Forschungsgemeinschaft (DFG, German Research Foundation), in part through the Collaborative Research Center [The Low-Energy Frontier of the Standard Model, Projektnummer 204404729 - SFB 1044], and in part through the Cluster of Excellence [Precision Physics, Fundamental Interactions, and Structure of Matter] (PRISMA$^+$ EXC 2118/1) within the German Excellence Strategy (Project ID 39083149).

\appendix
\section{Amplitudes}
\label{appA}
\subsection{Representations of the invariant amplitude}\label{app:Amp.rep}
We write the scattering amplitude $\mathcal{T}$  as 
\begin{align}
\mathcal{T} = &\bar u(p') \left[ a_N q \cdot \epsilon \V + a_E \Ve + a_K q \cdot \epsilon \Vk + a_{EK} \Vek \right] u(p), \label{eq:ampV}
\end{align}
where $u(p)$ and $\bar{u} (p')= u^\dagger (p') \gamma^0$ are the Dirac spinors corresponding to the initial and final nucleon states respectively, $\epsilon$ is the photon polarization vector, and $q$ is the 4-momentum of the outgoing pion, the coefficients $a_N$, $a_E$, $a_K$ and $a_{EK}$ are complex functions of the Mandelstam variables, while there are four operators defined as
\begin{equation}
\V=\gamma^5, \quad \Ve=\slashed \epsilon \gamma^5, \quad \Vk=\slashed k \gamma^5, \quad \Vek=\slashed \epsilon \slashed k \gamma^5,
\label{eq:Vbasis}
\end{equation}
where $k$ is the photon 4-momentum. There is another representation, commonly used, in terms of Lorentz invariant operators, $M_i$, where the scattering amplitude reads
\begin{align}
\mathcal{T} = \epsilon_\mu \mathcal{M}^\mu = &\epsilon_\mu \bar u(p')\left(
	\sum_{i=1}^4A_iM_i^\mu\right)u(p), \label{eq:ampM}
\intertext{with $\mathcal{M}^\mu$ the factorized hadronic current and}
\nn \epsilon\cdot M_1 =& \mathrm{i} \slk\sle\gamma_5,\\
 \nn \epsilon\cdot M_2 =&\mathrm{i}(p'\cdot\epsilon k\cdot q - q\cdot\epsilon k\cdot(p+p'))\gamma_5,\\
\nn \epsilon\cdot M_3 =&\mathrm{i}(\sle k\cdot q- \slk q\cdot\epsilon)\gamma_5,\\
 \nn \epsilon\cdot M_4 =&\mathrm{i}(\sle k \cdot(p+p')-\slk p'\cdot \epsilon - 2m_N \slk\sle)\gamma_5.
\intertext{Note that in the c.m. system $p\cdot\epsilon=0$. One can easily find the conversion between the two different representations:}
 A_1 = & 
\mathrm{i}\left(a_{EK}-\frac{m_N}{k\cdot p}\left(a_E+k\cdot q a_K\right)\right),\label{coefs:A1}\\
 A_2 = &
\mathrm{i}\frac{a_N}{2 k\cdot p},\\
 A_3 = &
\mathrm{i}\left(a_K\left(1-\frac{k\cdot q}{2k\cdot p}\right)-\frac{a_E}{2k\cdot p}\right),\\
 A_4 = &
-\frac{\mathrm{i}}{2k\cdot p}\left(a_E+k\cdot q a_K\right).
\label{coefs:A4}
\end{align}
For practical purposes, as explained in Sec.~\ref{subsec:kin.amp.obs}, it is sometimes convenient to use the CGLN amplitudes~\cite{Chew:1957tf}. In this way the scattering amplitude from Eq.~(\ref{eq:scatt.amp}) reads
\begin{align}
\nn \mathcal{T} = \frac{4\pi W}{m_N}\chi_f^\dagger\mathcal{F}\chi_i,
\end{align}
where $W=\sqrt{s}$ is the center-of-mass energy, and the amplitude $\mathcal{F}$ can be expressed as the decomposition of the $\mathcal{F}_i$ ($i=1,\ldots ,4$) pieces as shown in Eq.~(\ref{eq:Fdecomposition}). These pieces are given explicitly by
\begin{align}
\mathcal{F}_1 =& -i\frac{N_1 N_2}{8 \pi  W} (W-m_N) \left( A_1 + \frac{k\cdot q}{W-m_N} A_3+ \left(W-m_N-\frac{k\cdot q}{W-m_N} \right)A_4 \right), \label{coefs:F1} \\
\mathcal{F}_2 =& - i \frac{ |\vec{q^*}| N_1 N_2 }{8 \pi  W (E_2+m_N)} (W-m_N) \left( \frac{A_3 k\cdot q}{m_N+W}+ A_4 \left(-\frac{k\cdot q}{m_N+W}+m_N+W\right)- A_1
\right), \\
\mathcal{F}_3 =& - i \frac{ |\vec {q^*}| N_1 N_2 }{8 \pi  W} (W-m_N) ( A_2 (W-m_N)+ A_3 - A_4 ), \\
\mathcal{F}_4 =& - i \frac{ |\vec {q^*}|^2 N_1 N_2  }{8 \pi  W (E_2 +m_N)} (W-m_N) ( -A_2(m_N+W)+ A_3 - A_4 ) \label{coefs:F4},
\end{align}
with $N_i=\sqrt{m_N+E_i}$, $E_1=\sqrt{m_N^2+ |\vec {p}|^2}$, $E_2=\sqrt{m_N^2+ |\vec {q}|^2}$; $|\vec {p}|$ and $|\vec {q}|$ are evaluated in the c.m. system. $A_i$ are the coefficients of the scattering amplitude in the Lorentz invariant basis, $\{M_i\}$ as in Eq.~(\ref{eq:ampM}). Having the explicit expressions of $A_i$ in terms of the coefficients $a_N$, $a_E$, $a_K$ and $a_{EK}$ through the relations (\ref{coefs:A1})-(\ref{coefs:F4}), we are able to compute the observables as presented in Eqs.~(\ref{eq:dsigma}), (\ref{eq:Sigma}) and (\ref{eq:Tasym})  from the amplitude parametrized in the $\{V_N, V_E, V_K, V_{EK} \}$ basis of Eq.~(\ref{eq:Vbasis}). We write down the tree level amplitudes in this basis in the following Sec.~\ref{subsec:amp}.

\subsection{Tree level amplitude}\label{subsec:amp}

\subsubsection{At $O(p^1)$}\label{amp:order1}
\begin{eqnarray}
\mathcal{T}^{(1)}_{(a)}&=& C^{(1)}_I \frac{e g}{F_0} \Ve , \\
\mathcal{T}^{(1)}_{(b)}&=& C^{(1)}_{II} \frac{e g}{F_0} \left[\frac{ \left(s-m_N^2\right)}{ \left(m_2^2-s\right)} \Ve + \frac{ (m_N+m_2)}{ \left(m_2^2-s\right)} \Vek \right], \\
\mathcal{T}^{(1)}_{(c)}&=& C^{(1)}_{III} \frac{e g}{F_0} \left[\frac{ \left(m_N^2-u\right)}{ \left(m_2^2-u\right)} \Ve+\frac{2  (m_N+m_2)}{ \left(m_2^2-u\right)} q \cdot \epsilon \V + \frac{ (m_N+m_2)}{ \left(m_2^2-u\right)} \Vek \right], \\
\mathcal{T}^{(1)}_{(d)}&=& C^{(1)}_{IV} \frac{2 \sqrt{2} e g m_N }{F_0 \left(-2 m_N^2+s+u\right)} q \cdot \epsilon \V ,
\end{eqnarray}
where the coefficients $C_{i}^{(1)}$ for $i=\{I,II,III,IV\}$ are given in Table~\ref{coefs1} and $u\equiv(p-q)^2$. Here $m_N$ is the physical nucleon mass coming from the external legs in the Feynman diagrams of Fig.~\ref{fg:feynmanTree}, that in our case corresponds to the order $O(p^3)$ nucleon mass,  whose expression is derived in Eq.~(\ref{eq:m3}).  The inner nucleon propagator has the second order nucleon mass $m_2$ instead of $m$. This automatically generates the $O(p^2)$ and higher order contributions corresponding to $c_1$ mass insertions.

\begin{table}[h] \centering
\begin{tabular}
[c]{|l|c|c|c|c|}\hline
Channel & $C_{I}^{(1)}$ & $C_{II}^{(1)}$ & $C_{III}^{(1)}$ & $C_{IV}^{(1)}$ \\ \hline
$\gamma p \rightarrow p \pi^0$ & $0$                   & $\frac{1}{2}$        & $\frac{1}{2}$ & $0$ \\
$\gamma p \rightarrow n \pi^+$ & $\frac{1}{\sqrt{2}}$  & $\frac{1}{\sqrt{2}}$ & $0$ & $-1$ \\
$\gamma n \rightarrow p \pi^-$ & $-\frac{1}{\sqrt{2}}$ & $0$                  & $\frac{1}{\sqrt{2}}$ & $1$ \\
$\gamma n \rightarrow n \pi^0$ & $0$                   & $0$                  & $0$ & $0$ \\ \hline
\end{tabular}
\caption{Tree level amplitude constants for each channel at order $\mathcal{O}(q^1)$.
\label{coefs1}}
\end{table}

\subsubsection{At $O(p^2)$}\label{amp:order2}

In what follows, for the amplitudes of order $O(p^2)$ and higher, the leading order bare constants, e.g. $m$, can be replaced by their corresponding physical ones since the difference  is of higher order than  our current accuracy [$O(p^3)$]. This replacement is actually made in our calculation: 
\begin{eqnarray}
\mathcal{T}^{(2)}_{(b)}&=& C^{(2)}_{II} \frac{e g_A}{F} \left[\frac{ \left(3 m^2+s\right)}{ 2m \left( m^2-  s\right)} \Vek -\Ve \right], \\
\mathcal{T}^{(2)}_{(c)}&=& C^{(2)}_{III} \frac{e g_A}{F} \left[\frac{2 }{ \left(m^2-u\right)} q \cdot \epsilon \Vk + \frac{ \left(3 m^2+u\right)}{ 2m \left( m^2- u\right)} \Vek + \Ve \right].
\end{eqnarray}
The definitions for the constants $C_{II}^{(2)}$ and $C_{III}^{(2)}$ are presented in Table~\ref{coefs2}.

\begin{table}[h] \centering
\begin{tabular}
[c]{|l|c|c|c|c|}\hline
Channel & $C_{I}^{(2)}$ & $C_{II}^{(2)}$ & $C_{III}^{(2)}$ & $C_{IV}^{(2)}$ \\ \hline
$\gamma p \rightarrow p \pi^0$&$0$ & $\frac{1}{2}(c_6 + c_7)$& $\frac{1}{2}(c_6 + c_7)$ & $0$ \\
$\gamma p \rightarrow n \pi^+$&$0$ & $\frac{1}{\sqrt{2}}(c_6 + c_7)$& $\frac{1}{\sqrt{2}}c_7$ & $0$  \\
$\gamma n \rightarrow p \pi^-$&$0$ & $\frac{1}{\sqrt{2}}c_7$ & $\frac{1}{\sqrt{2}}(c_6 + c_7)$ & $0$ \\
$\gamma n \rightarrow n \pi^0$&$0$ & $-\frac{1}{2}c_7$       & $-\frac{1}{2}c_7$          & $0$ \\ \hline
\end{tabular}
\caption{Tree level amplitude constants for each channel at order $\mathcal{O}(q^2)$.  \label{coefs2}}
\end{table}

\subsubsection{At $O(p^3)$}\label{amp:order3}

\begin{eqnarray}
\mathcal{T}^{(3)}_{(a)}&=& C^{(3)}_{Ia} \frac{e}{F} \left[\frac{2 \left(2 m^2-m_\pi^2+s+u\right)}{ m} \Vek + \frac{2 \left(s-m^2\right)}{ m} q \cdot \epsilon \V + 4 q \cdot \epsilon \Vk+ 2 (s-u) \Ve \right] \nonumber \\
&+& C^{(3)}_{Ib} \frac{e}{F} \left[ \left( \frac{ m_{\pi }^2 \left(2 m^2-s-u\right)+2 \left(m^2-s\right) \left(m^2-u\right)}{4 \sqrt{2}  m^2} d_{20} + \frac{ \left(2 m^2-s-u\right)d_{212} -4 m_{\pi }^2 d_{168}}{2 \sqrt{2} }\right) \Ve \right. \nn \\
&& \hspace{2.5cm} \left. + \left(\frac{ \left(m^2+m_{\pi }^2-s\right)}{2 \sqrt{2} m^2}  d_{20}+\frac{ d_{212}}{\sqrt{2} }\right) q \cdot \epsilon \Vk \right], \\
\mathcal{T}^{(3)}_{(b)}&=& C^{(3)}_{II} \frac{e}{F} d_{168} \left[\frac{2 m\, m_\pi^2 }{ m^2- s} \Vek -  m_\pi^2 \Ve \right], \\
\mathcal{T}^{(3)}_{(c)}&=& C^{(3)}_{III} \frac{e}{F} d_{168} \left[\frac{4  m\, m_\pi^2 }{ m^2- u} q \cdot \epsilon \V +\frac{2  m\, m_\pi^2}{ m^2- u} \Vek+  m_\pi^2 \Ve\right], \\
\mathcal{T}^{(3)}_{(d)}&=& C^{(3)}_{IV} \frac{e}{F} d_{168} \frac{4 \sqrt{2}  m\, m_\pi^2 }{ \left(2 m^2-s-u\right)}q \cdot \epsilon \V,
\end{eqnarray}
where $d_{168}=2d_{16}-d_{18}$ and $d_{212}=2 d_{21}-d_{22}$. The coefficients $C_{j}^{(3)}$ for $j=\{ Ia,Ib,II,III,IV \}$ are given in Table~\ref{coefs3}.

\begin{table}[h] \centering
\begin{tabular}
[c]{|l|c|c|c|c|c|}\hline
Channel &$C_{Ia}^{(3)}$ & $C_{Ib}^{(3)}$ & $C_{II}^{(3)}$ & $C_{III}^{(3)}$ & $C_{IV}^{(3)}$ \\ \hline
$\gamma p \rightarrow p \pi^0$&$d_8+d_9$ & $0$   & $1$   & $1$ & $0$ \\
$\gamma p \rightarrow n \pi^+$&$\sqrt{2}d_9$ & $-1$  & $\sqrt{2}$ & $0$ & $1$ \\
$\gamma n \rightarrow p \pi^-$&$\sqrt{2}d_9$ & $1$ & $0$    & $\sqrt{2}$ & $-1$ \\
$\gamma n \rightarrow n \pi^0$&$d_8-d_9$ & $0$    & $0$     & $0$ & $0$ \\ \hline
\end{tabular}
\caption{Tree level amplitude constants for each channel at order $\mathcal{O}(q^3)$. \label{coefs3}}
\end{table}

\subsubsection{At $O(p^{5/2})$}\label{amp:order5/2}

For the following amplitudes the definitions of the constants $D_{II}$ and $D_{III}$ are given in Table~\ref{coefs25}.
\begin{align}
&\mathcal{T}^{(5/2)}_{(a)} = D_{II} \frac{e g_M h_A}{4 F m\,m_\Delta (m+m_\Delta)} \left[\frac{ \left(m^2-s\right) (m \,m_\Delta+s)}{ \left(-i \Gamma_\Delta(s) m_\Delta+m_\Delta^2-s\right)}q \cdot \epsilon \V -\frac{ \left(m^2 m_\Delta+2 m s+m_\Delta s\right)}{ \left(-i \Gamma_\Delta(s) m_\Delta+m_\Delta^2-s\right)}q \cdot \epsilon \Vk \right. \nonumber \\
+ & \frac{ \left(m^4-8 m^3 m_\Delta-m^2 \left(m_\pi^2+6 s\right)+4 m\, m_\Delta \left(m_\pi^2-2 s\right)+s \left(5 m_\pi^2-5 s-6 u\right)\right)}{6 \left(-i \Gamma_\Delta(s) m_\Delta+m_\Delta^2-s\right)}\Vek \nonumber \\
+ &\left. \frac{ \left(-m^5+3 m^4 m_\Delta+m^3 \left(m_\pi^2-2 s\right)+m^2 m_\Delta \left(m_\pi^2-10 s\right)-m s \left(m_\pi^2+3 s-6 u\right)+m_\Delta s \left(-m_\pi^2+s+6 u\right)\right)}{6 \left(-i \Gamma_\Delta(s) m_\Delta+m_\Delta^2-s\right)} \Ve \right],
\end{align}

\begin{align}
&\mathcal{T}^{(5/2)}_{(b)}= D_{III} \frac{e g_M h_A}{4 F m \,m_\Delta (m+m_\Delta)} \left[-\frac{ \left(m^2-s\right) (m\, m_\Delta+u)}{ \left(m_\Delta^2-u\right)}q \cdot \epsilon \V \right.\nonumber \\ 
+&\frac{ \left(-m^3+6 m^2 m_\Delta+m \left(m_\pi^2+3 u\right)+m_\Delta \left(m_\pi^2-3 s-u\right)\right)}{3 \left(m_\Delta^2-u\right)}q \cdot \epsilon \Vk \nonumber \\ 
+&  \frac{ \left(-m^4+8 m^3 m_\Delta+m^2 \left(m_\pi^2+6 u\right)-4 m\, m_\Delta \left(m_\pi^2-2 u\right)+u \left(-5 m_\pi^2+6 s+5 u\right)\right)}{6 \left(m_\Delta^2-u\right)} \Vek \nonumber \\
+& \left. \frac{ \left(-m^5+3 m^4 m_\Delta+m^3 \left(m_\pi^2-2 u\right)+m^2 m_\Delta \left(m_\pi^2-10 u\right)-m u \left(m_\pi^2-6 s+3 u\right)+m_\Delta u \left(-m_\pi^2+6 s+u\right)\right)}{6 \left(m_\Delta^2-u\right)} \Ve \right],
\end{align}
where the energy dependent width, $\Gamma_{\Delta}(s)$, is given by \cite{Gegelia:2016pjm} 
\begin{align}
    \Gamma_{\Delta}(s)=\frac{\left( h_A/2\right)^2 \Lambda^{3/2}(s, m_{\pi}^2,m^2)}{192 \pi F^2 s^3} \left[ \left(s-m_{\pi}^2 + m^2 \right)m_\Delta + 2 s m \right] \theta \left( s-(m+m_{\pi})^2 \right),
\end{align}
using $\theta(x)$ as the step function ensuring the dependence to be above the threshold of pion production on nucleons.

\begin{table}[h] \centering
\begin{tabular}
[c]{|l|c|c|}\hline
Channel & $D_{II}$ & $D_{III}$ \\ \hline
$\gamma p \rightarrow p \pi^0$ & $1$& $-1$  \\
$\gamma p \rightarrow n \pi^+$ & $\frac{-1}{\sqrt{2}}$& $\frac{-1}{\sqrt{2}}$   \\
$\gamma n \rightarrow p \pi^-$ &  $\frac{1}{\sqrt{2}}$ &  $\frac{1}{\sqrt{2}}$  \\
$\gamma n \rightarrow n \pi^0$ & $1$ & $-1$  \\ \hline
\end{tabular}
\caption{Tree level amplitude constants for each channel at order $\mathcal{O}(q^{5/2})$. \label{coefs25}}
\end{table}

\section{Renormalization factors}
\label{AppB}

 The wave function renormalization of the external legs is written as
\begin{align}
\mathcal{Z}_N= &1 + \delta^{(2)}_{\mathcal{Z}_N} + \mathcal{O}(p^3), \qquad \mathcal{Z}^{(2)}_\pi=1 + \delta^{(2)}_{\mathcal{Z}_\pi} + \mathcal{O}(p^3),\\
\intertext{ where}
\delta^{(2)}_{\mathcal{Z}_N} = &-\frac{3 g_A^2}{64 \pi ^2 F^2 \left(m_\pi^2-4 m^2\right)} \Bigg \{ 4 m_\pi^2 \left(A_0\left[m^2\right]+\left(m_\pi^2-3 m^2\right) B_0\left[m^2,m_\pi^2,m^2\right]-m^2\right) \nonumber \\
&+ \left(12 m^2-5 m_\pi^2\right) A_0\left[m_\pi^2\right] \Bigg \}, \nonumber \\
\delta^{(2)}_{\mathcal{Z}_\pi} = & -\frac{2}{3 F^2} \left\lbrace 3 l_4 m_\pi^2 + \frac{A_0\left[m_\pi^2\right]}{16 \pi ^2 }\right\rbrace .\\
\intertext{ Furthermore, the mass corrections are given by}
m_N=&\tilde{m} -4 \tilde{c}_1  m_\pi^2 + \delta^{(3)}_{m} + \mathcal{O}\left( p^4 \right),
\label{eq:m3}\\
m_2 = & \tilde m -4 \tilde c_1 m_\pi^2 = m_N - \delta^{(3)}_{m}+ \mathcal{O}\left( p^4 \right),
\label{eq:m2}\\
\intertext{with}
\delta^{(3)}_{m}=& \frac{3 g_A^2 m\, m_\pi^2}{32 \pi ^2 F^2} \left\lbrace \bar{B}_0\left[m^2,m_\pi^2,m^2\right] -\left(1+\frac{\bar{A}_0\left[m^2\right]}{m^2} \right) \right\rbrace.\\
\intertext{
Finally, the corrections to the axial vector coupling and the pion decay constant read}
g_A = &\tilde g \left( 1 + \frac{ 4 d^r_{16} m^2_\pi}{g_A} + \delta^{(2)}_{g_A} \right) + \mathcal{O}(p^3), \qquad  F=F_0 \left( 1+ \delta^{(2)}_F \right) + \mathcal{O}(p^3),\\
\intertext{where}
\delta^{(2)}_{g_A} =& \frac{1}{16 \pi ^2 F^2 \left(4 m^2-m_\pi^2\right)} \Bigg \{ 4 g_A^2 m_\pi^2 \bar{A}_0\left[m^2\right]+\left(\left(8 g_A^2+4\right) m^2-\left(4 g_A^2+1\right) m_\pi^2\right) \bar{A}_0\left[m_\pi^2\right] \nonumber \\
& + m_\pi^2 \left(\left(\left(3 g_A^2+2\right) m_\pi^2-8 \left(g_A^2+1\right) m^2\right) \bar{B}_0\left[m^2,m_\pi^2,m^2\right]-4 g_A^2 m^2\right) \Bigg \}, \nonumber \\
\delta^{(2)}_F =& \frac{l_4^r m_\pi^2}{F^2} + \frac{\bar{A}_0\left[m_\pi^2\right]}{16 \pi ^2 F^2}.
\end{align}

 Note here that $l_4^r$ and $d_{16}^r$ are $\widetilde{\rm MS}$-renormalized LECs.

\bibliography{weak}

\begin{thebibliography}{89}
\expandafter\ifx\csname natexlab\endcsname\relax\def\natexlab#1{#1}\fi
\expandafter\ifx\csname bibnamefont\endcsname\relax
  \def\bibnamefont#1{#1}\fi
\expandafter\ifx\csname bibfnamefont\endcsname\relax
  \def\bibfnamefont#1{#1}\fi
\expandafter\ifx\csname citenamefont\endcsname\relax
  \def\citenamefont#1{#1}\fi
\expandafter\ifx\csname url\endcsname\relax
  \def\url#1{\texttt{#1}}\fi
\expandafter\ifx\csname urlprefix\endcsname\relax\def\urlprefix{URL }\fi
\providecommand{\bibinfo}[2]{#2}
\providecommand{\eprint}[2][]{\url{#2}}

\bibitem[{\citenamefont{Adler}(1968)}]{Adler:1968tw}
\bibinfo{author}{\bibfnamefont{S.~L.} \bibnamefont{Adler}},
  \bibinfo{journal}{Annals Phys.} \textbf{\bibinfo{volume}{50}},
  \bibinfo{pages}{189} (\bibinfo{year}{1968}), \bibinfo{note}{[,225(1968)]}.

\bibitem[{\citenamefont{Drechsel et~al.}(2007)\citenamefont{Drechsel, Kamalov,
  and Tiator}}]{Drechsel:2007if}
\bibinfo{author}{\bibfnamefont{D.}~\bibnamefont{Drechsel}},
  \bibinfo{author}{\bibfnamefont{S.~S.} \bibnamefont{Kamalov}},
  \bibnamefont{and} \bibinfo{author}{\bibfnamefont{L.}~\bibnamefont{Tiator}},
  \bibinfo{journal}{Eur. Phys. J.} \textbf{\bibinfo{volume}{A34}},
  \bibinfo{pages}{69} (\bibinfo{year}{2007}), \eprint{0710.0306}.

\bibitem[{\citenamefont{Weinberg}(1979)}]{Weinberg:1978kz}
\bibinfo{author}{\bibfnamefont{S.}~\bibnamefont{Weinberg}},
  \bibinfo{journal}{Physica} \textbf{\bibinfo{volume}{A96}},
  \bibinfo{pages}{327} (\bibinfo{year}{1979}).

\bibitem[{\citenamefont{Gasser and Leutwyler}(1984)}]{Gasser:1983yg}
\bibinfo{author}{\bibfnamefont{J.}~\bibnamefont{Gasser}} \bibnamefont{and}
  \bibinfo{author}{\bibfnamefont{H.}~\bibnamefont{Leutwyler}},
  \bibinfo{journal}{Annals Phys.} \textbf{\bibinfo{volume}{158}},
  \bibinfo{pages}{142} (\bibinfo{year}{1984}).

\bibitem[{\citenamefont{Gasser and Leutwyler}(1985)}]{Gasser:1984gg}
\bibinfo{author}{\bibfnamefont{J.}~\bibnamefont{Gasser}} \bibnamefont{and}
  \bibinfo{author}{\bibfnamefont{H.}~\bibnamefont{Leutwyler}},
  \bibinfo{journal}{Nucl. Phys.} \textbf{\bibinfo{volume}{B250}},
  \bibinfo{pages}{465} (\bibinfo{year}{1985}).

\bibitem[{\citenamefont{Scherer and Schindler}(2012)}]{Scherer:2012xha}
\bibinfo{author}{\bibfnamefont{S.}~\bibnamefont{Scherer}} \bibnamefont{and}
  \bibinfo{author}{\bibfnamefont{M.~R.} \bibnamefont{Schindler}},
  \bibinfo{journal}{Lect. Notes Phys.} \textbf{\bibinfo{volume}{830}},
  \bibinfo{pages}{pp.1} (\bibinfo{year}{2012}).

\bibitem[{\citenamefont{Kroll and Ruderman}(1954)}]{PhysRev.93.233}
\bibinfo{author}{\bibfnamefont{N.~M.} \bibnamefont{Kroll}} \bibnamefont{and}
  \bibinfo{author}{\bibfnamefont{M.~A.} \bibnamefont{Ruderman}},
  \bibinfo{journal}{Phys. Rev.} \textbf{\bibinfo{volume}{93}},
  \bibinfo{pages}{233} (\bibinfo{year}{1954}),
  \urlprefix\url{https://link.aps.org/doi/10.1103/PhysRev.93.233}.

\bibitem[{\citenamefont{De~Baenst}(1970)}]{DeBaenst:1971hp}
\bibinfo{author}{\bibfnamefont{P.}~\bibnamefont{De~Baenst}},
  \bibinfo{journal}{Nucl. Phys.} \textbf{\bibinfo{volume}{B24}},
  \bibinfo{pages}{633} (\bibinfo{year}{1970}).

\bibitem[{\citenamefont{Vainshtein and Zakharov}(1972)}]{Vainshtein:1972ih}
\bibinfo{author}{\bibfnamefont{A.~I.} \bibnamefont{Vainshtein}}
  \bibnamefont{and} \bibinfo{author}{\bibfnamefont{V.~I.}
  \bibnamefont{Zakharov}}, \bibinfo{journal}{Nucl. Phys.}
  \textbf{\bibinfo{volume}{B36}}, \bibinfo{pages}{589} (\bibinfo{year}{1972}).

\bibitem[{\citenamefont{Mazzucato et~al.}(1986)}]{Mazzucato:1986dz}
\bibinfo{author}{\bibfnamefont{E.}~\bibnamefont{Mazzucato}}
  \bibnamefont{et~al.}, \bibinfo{journal}{Phys. Rev. Lett.}
  \textbf{\bibinfo{volume}{57}}, \bibinfo{pages}{3144} (\bibinfo{year}{1986}).

\bibitem[{\citenamefont{Beck et~al.}(1990)\citenamefont{Beck, Kalleicher,
  Schoch, Vogt, Koch, Stroher, Metag, McGeorge, Kellie, and
  Hall}}]{Beck:1990da}
\bibinfo{author}{\bibfnamefont{R.}~\bibnamefont{Beck}},
  \bibinfo{author}{\bibfnamefont{F.}~\bibnamefont{Kalleicher}},
  \bibinfo{author}{\bibfnamefont{B.}~\bibnamefont{Schoch}},
  \bibinfo{author}{\bibfnamefont{J.}~\bibnamefont{Vogt}},
  \bibinfo{author}{\bibfnamefont{G.}~\bibnamefont{Koch}},
  \bibinfo{author}{\bibfnamefont{H.}~\bibnamefont{Stroher}},
  \bibinfo{author}{\bibfnamefont{V.}~\bibnamefont{Metag}},
  \bibinfo{author}{\bibfnamefont{J.~C.} \bibnamefont{McGeorge}},
  \bibinfo{author}{\bibfnamefont{J.~D.} \bibnamefont{Kellie}},
  \bibnamefont{and} \bibinfo{author}{\bibfnamefont{S.~J.} \bibnamefont{Hall}},
  \bibinfo{journal}{Phys. Rev. Lett.} \textbf{\bibinfo{volume}{65}},
  \bibinfo{pages}{1841} (\bibinfo{year}{1990}).

\bibitem[{\citenamefont{Drechsel and Tiator}(1992)}]{Drechsel:1992pn}
\bibinfo{author}{\bibfnamefont{D.}~\bibnamefont{Drechsel}} \bibnamefont{and}
  \bibinfo{author}{\bibfnamefont{L.}~\bibnamefont{Tiator}},
  \bibinfo{journal}{J. Phys.} \textbf{\bibinfo{volume}{G18}},
  \bibinfo{pages}{449} (\bibinfo{year}{1992}).

\bibitem[{\citenamefont{Bernard and Meissner}(2007)}]{Bernard:2006gx}
\bibinfo{author}{\bibfnamefont{V.}~\bibnamefont{Bernard}} \bibnamefont{and}
  \bibinfo{author}{\bibfnamefont{U.-G.} \bibnamefont{Meissner}},
  \bibinfo{journal}{Ann. Rev. Nucl. Part. Sci.} \textbf{\bibinfo{volume}{57}},
  \bibinfo{pages}{33} (\bibinfo{year}{2007}), \eprint{hep-ph/0611231}.

\bibitem[{\citenamefont{Bernard et~al.}(1991)\citenamefont{Bernard, Kaiser,
  Gasser, and Meissner}}]{Bernard:1991rt}
\bibinfo{author}{\bibfnamefont{V.}~\bibnamefont{Bernard}},
  \bibinfo{author}{\bibfnamefont{N.}~\bibnamefont{Kaiser}},
  \bibinfo{author}{\bibfnamefont{J.}~\bibnamefont{Gasser}}, \bibnamefont{and}
  \bibinfo{author}{\bibfnamefont{U.~G.} \bibnamefont{Meissner}},
  \bibinfo{journal}{Phys. Lett.} \textbf{\bibinfo{volume}{B268}},
  \bibinfo{pages}{291} (\bibinfo{year}{1991}).

\bibitem[{\citenamefont{Bernard et~al.}(1992)\citenamefont{Bernard, Kaiser, and
  Meissner}}]{Bernard:1992nc}
\bibinfo{author}{\bibfnamefont{V.}~\bibnamefont{Bernard}},
  \bibinfo{author}{\bibfnamefont{N.}~\bibnamefont{Kaiser}}, \bibnamefont{and}
  \bibinfo{author}{\bibfnamefont{U.~G.} \bibnamefont{Meissner}},
  \bibinfo{journal}{Nucl. Phys.} \textbf{\bibinfo{volume}{B383}},
  \bibinfo{pages}{442} (\bibinfo{year}{1992}).

\bibitem[{\citenamefont{Bernard et~al.}(2001)\citenamefont{Bernard, Kaiser, and
  Meissner}}]{Bernard:2001gz}
\bibinfo{author}{\bibfnamefont{V.}~\bibnamefont{Bernard}},
  \bibinfo{author}{\bibfnamefont{N.}~\bibnamefont{Kaiser}}, \bibnamefont{and}
  \bibinfo{author}{\bibfnamefont{U.-G.} \bibnamefont{Meissner}},
  \bibinfo{journal}{Eur. Phys. J.} \textbf{\bibinfo{volume}{A11}},
  \bibinfo{pages}{209} (\bibinfo{year}{2001}), \eprint{hep-ph/0102066}.

\bibitem[{\citenamefont{Hornidge et~al.}(2013)}]{Hornidge:2012ca}
\bibinfo{author}{\bibfnamefont{D.}~\bibnamefont{Hornidge}} \bibnamefont{et~al.}
  (\bibinfo{collaboration}{A2, CB-TAPS}), \bibinfo{journal}{Phys. Rev. Lett.}
  \textbf{\bibinfo{volume}{111}}, \bibinfo{pages}{062004}
  (\bibinfo{year}{2013}), \eprint{1211.5495}.

\bibitem[{\citenamefont{Fernandez-Ramirez and
  Bernstein}(2013)}]{FernandezRamirez:2012nw}
\bibinfo{author}{\bibfnamefont{C.}~\bibnamefont{Fernandez-Ramirez}}
  \bibnamefont{and} \bibinfo{author}{\bibfnamefont{A.~M.}
  \bibnamefont{Bernstein}}, \bibinfo{journal}{Phys. Lett.}
  \textbf{\bibinfo{volume}{B724}}, \bibinfo{pages}{253} (\bibinfo{year}{2013}),
  \eprint{1212.3237}.

\bibitem[{\citenamefont{Hilt et~al.}(2013{\natexlab{a}})\citenamefont{Hilt,
  Scherer, and Tiator}}]{Hilt:2013uf}
\bibinfo{author}{\bibfnamefont{M.}~\bibnamefont{Hilt}},
  \bibinfo{author}{\bibfnamefont{S.}~\bibnamefont{Scherer}}, \bibnamefont{and}
  \bibinfo{author}{\bibfnamefont{L.}~\bibnamefont{Tiator}},
  \bibinfo{journal}{Phys. Rev.} \textbf{\bibinfo{volume}{C87}},
  \bibinfo{pages}{045204} (\bibinfo{year}{2013}{\natexlab{a}}),
  \eprint{1301.5576}.

\bibitem[{\citenamefont{Bijnens and Ecker}(2014)}]{Bijnens:2014lea}
\bibinfo{author}{\bibfnamefont{J.}~\bibnamefont{Bijnens}} \bibnamefont{and}
  \bibinfo{author}{\bibfnamefont{G.}~\bibnamefont{Ecker}},
  \bibinfo{journal}{Ann. Rev. Nucl. Part. Sci.} \textbf{\bibinfo{volume}{64}},
  \bibinfo{pages}{149} (\bibinfo{year}{2014}), \eprint{1405.6488}.

\bibitem[{\citenamefont{Bernard}(2008)}]{Bernard:2007zu}
\bibinfo{author}{\bibfnamefont{V.}~\bibnamefont{Bernard}},
  \bibinfo{journal}{Prog. Part. Nucl. Phys.} \textbf{\bibinfo{volume}{60}},
  \bibinfo{pages}{82} (\bibinfo{year}{2008}), \eprint{0706.0312}.

\bibitem[{\citenamefont{Gasser et~al.}(1988)\citenamefont{Gasser, Sainio, and
  Svarc}}]{Gasser:1987rb}
\bibinfo{author}{\bibfnamefont{J.}~\bibnamefont{Gasser}},
  \bibinfo{author}{\bibfnamefont{M.~E.} \bibnamefont{Sainio}},
  \bibnamefont{and} \bibinfo{author}{\bibfnamefont{A.}~\bibnamefont{Svarc}},
  \bibinfo{journal}{Nucl. Phys.} \textbf{\bibinfo{volume}{B307}},
  \bibinfo{pages}{779} (\bibinfo{year}{1988}).

\bibitem[{\citenamefont{Jenkins and
  Manohar}(1991{\natexlab{a}})}]{Jenkins:1990jv}
\bibinfo{author}{\bibfnamefont{E.~E.} \bibnamefont{Jenkins}} \bibnamefont{and}
  \bibinfo{author}{\bibfnamefont{A.~V.} \bibnamefont{Manohar}},
  \bibinfo{journal}{Phys. Lett.} \textbf{\bibinfo{volume}{B255}},
  \bibinfo{pages}{558} (\bibinfo{year}{1991}{\natexlab{a}}).

\bibitem[{\citenamefont{Jenkins and
  Manohar}(1991{\natexlab{b}})}]{Jenkins:1991es}
\bibinfo{author}{\bibfnamefont{E.~E.} \bibnamefont{Jenkins}} \bibnamefont{and}
  \bibinfo{author}{\bibfnamefont{A.~V.} \bibnamefont{Manohar}},
  \bibinfo{journal}{Phys. Lett.} \textbf{\bibinfo{volume}{B259}},
  \bibinfo{pages}{353} (\bibinfo{year}{1991}{\natexlab{b}}).

\bibitem[{\citenamefont{Becher and Leutwyler}(1999)}]{Becher:1999he}
\bibinfo{author}{\bibfnamefont{T.}~\bibnamefont{Becher}} \bibnamefont{and}
  \bibinfo{author}{\bibfnamefont{H.}~\bibnamefont{Leutwyler}},
  \bibinfo{journal}{Eur. Phys. J.} \textbf{\bibinfo{volume}{C9}},
  \bibinfo{pages}{643} (\bibinfo{year}{1999}), \eprint{hep-ph/9901384}.

\bibitem[{\citenamefont{Fuchs et~al.}(2003)\citenamefont{Fuchs, Gegelia,
  Japaridze, and Scherer}}]{Fuchs:2003qc}
\bibinfo{author}{\bibfnamefont{T.}~\bibnamefont{Fuchs}},
  \bibinfo{author}{\bibfnamefont{J.}~\bibnamefont{Gegelia}},
  \bibinfo{author}{\bibfnamefont{G.}~\bibnamefont{Japaridze}},
  \bibnamefont{and} \bibinfo{author}{\bibfnamefont{S.}~\bibnamefont{Scherer}},
  \bibinfo{journal}{Phys. Rev.} \textbf{\bibinfo{volume}{D68}},
  \bibinfo{pages}{056005} (\bibinfo{year}{2003}), \eprint{hep-ph/0302117}.

\bibitem[{\citenamefont{Alarcon et~al.}(2013)\citenamefont{Alarcon,
  Martin~Camalich, and Oller}}]{Alarcon:2012kn}
\bibinfo{author}{\bibfnamefont{J.~M.} \bibnamefont{Alarcon}},
  \bibinfo{author}{\bibfnamefont{J.}~\bibnamefont{Martin~Camalich}},
  \bibnamefont{and} \bibinfo{author}{\bibfnamefont{J.~A.} \bibnamefont{Oller}},
  \bibinfo{journal}{Annals Phys.} \textbf{\bibinfo{volume}{336}},
  \bibinfo{pages}{413} (\bibinfo{year}{2013}), \eprint{1210.4450}.

\bibitem[{\citenamefont{Geng et~al.}(2008)\citenamefont{Geng, Martin~Camalich,
  Alvarez-Ruso, and Vicente~Vacas}}]{Geng:2008mf}
\bibinfo{author}{\bibfnamefont{L.~S.} \bibnamefont{Geng}},
  \bibinfo{author}{\bibfnamefont{J.}~\bibnamefont{Martin~Camalich}},
  \bibinfo{author}{\bibfnamefont{L.}~\bibnamefont{Alvarez-Ruso}},
  \bibnamefont{and} \bibinfo{author}{\bibfnamefont{M.~J.}
  \bibnamefont{Vicente~Vacas}}, \bibinfo{journal}{Phys. Rev. Lett.}
  \textbf{\bibinfo{volume}{101}}, \bibinfo{pages}{222002}
  (\bibinfo{year}{2008}), \eprint{0805.1419}.

\bibitem[{\citenamefont{Martin~Camalich
  et~al.}(2010)\citenamefont{Martin~Camalich, Geng, and
  Vicente~Vacas}}]{MartinCamalich:2010fp}
\bibinfo{author}{\bibfnamefont{J.}~\bibnamefont{Martin~Camalich}},
  \bibinfo{author}{\bibfnamefont{L.~S.} \bibnamefont{Geng}}, \bibnamefont{and}
  \bibinfo{author}{\bibfnamefont{M.~J.} \bibnamefont{Vicente~Vacas}},
  \bibinfo{journal}{Phys. Rev.} \textbf{\bibinfo{volume}{D82}},
  \bibinfo{pages}{074504} (\bibinfo{year}{2010}), \eprint{1003.1929}.

\bibitem[{\citenamefont{Fuchs et~al.}(2004)\citenamefont{Fuchs, Gegelia, and
  Scherer}}]{Fuchs:2003ir}
\bibinfo{author}{\bibfnamefont{T.}~\bibnamefont{Fuchs}},
  \bibinfo{author}{\bibfnamefont{J.}~\bibnamefont{Gegelia}}, \bibnamefont{and}
  \bibinfo{author}{\bibfnamefont{S.}~\bibnamefont{Scherer}},
  \bibinfo{journal}{J. Phys.} \textbf{\bibinfo{volume}{G30}},
  \bibinfo{pages}{1407} (\bibinfo{year}{2004}), \eprint{nucl-th/0305070}.

\bibitem[{\citenamefont{Lehnhart et~al.}(2005)\citenamefont{Lehnhart, Gegelia,
  and Scherer}}]{Lehnhart:2004vi}
\bibinfo{author}{\bibfnamefont{B.~C.} \bibnamefont{Lehnhart}},
  \bibinfo{author}{\bibfnamefont{J.}~\bibnamefont{Gegelia}}, \bibnamefont{and}
  \bibinfo{author}{\bibfnamefont{S.}~\bibnamefont{Scherer}},
  \bibinfo{journal}{J. Phys.} \textbf{\bibinfo{volume}{G31}},
  \bibinfo{pages}{89} (\bibinfo{year}{2005}), \eprint{hep-ph/0412092}.

\bibitem[{\citenamefont{Schindler
  et~al.}(2007{\natexlab{a}})\citenamefont{Schindler, Fuchs, Gegelia, and
  Scherer}}]{Schindler:2006it}
\bibinfo{author}{\bibfnamefont{M.~R.} \bibnamefont{Schindler}},
  \bibinfo{author}{\bibfnamefont{T.}~\bibnamefont{Fuchs}},
  \bibinfo{author}{\bibfnamefont{J.}~\bibnamefont{Gegelia}}, \bibnamefont{and}
  \bibinfo{author}{\bibfnamefont{S.}~\bibnamefont{Scherer}},
  \bibinfo{journal}{Phys. Rev.} \textbf{\bibinfo{volume}{C75}},
  \bibinfo{pages}{025202} (\bibinfo{year}{2007}{\natexlab{a}}),
  \eprint{nucl-th/0611083}.

\bibitem[{\citenamefont{Schindler
  et~al.}(2007{\natexlab{b}})\citenamefont{Schindler, Djukanovic, Gegelia, and
  Scherer}}]{Schindler:2006ha}
\bibinfo{author}{\bibfnamefont{M.~R.} \bibnamefont{Schindler}},
  \bibinfo{author}{\bibfnamefont{D.}~\bibnamefont{Djukanovic}},
  \bibinfo{author}{\bibfnamefont{J.}~\bibnamefont{Gegelia}}, \bibnamefont{and}
  \bibinfo{author}{\bibfnamefont{S.}~\bibnamefont{Scherer}},
  \bibinfo{journal}{Phys. Lett.} \textbf{\bibinfo{volume}{B649}},
  \bibinfo{pages}{390} (\bibinfo{year}{2007}{\natexlab{b}}),
  \eprint{hep-ph/0612164}.

\bibitem[{\citenamefont{Geng et~al.}(2009)\citenamefont{Geng, Martin~Camalich,
  and Vicente~Vacas}}]{Geng:2009ik}
\bibinfo{author}{\bibfnamefont{L.~S.} \bibnamefont{Geng}},
  \bibinfo{author}{\bibfnamefont{J.}~\bibnamefont{Martin~Camalich}},
  \bibnamefont{and} \bibinfo{author}{\bibfnamefont{M.~J.}
  \bibnamefont{Vicente~Vacas}}, \bibinfo{journal}{Phys. Rev.}
  \textbf{\bibinfo{volume}{D79}}, \bibinfo{pages}{094022}
  (\bibinfo{year}{2009}), \eprint{0903.4869}.

\bibitem[{\citenamefont{Alarcon et~al.}(2012)\citenamefont{Alarcon,
  Martin~Camalich, and Oller}}]{Alarcon:2011zs}
\bibinfo{author}{\bibfnamefont{J.~M.} \bibnamefont{Alarcon}},
  \bibinfo{author}{\bibfnamefont{J.}~\bibnamefont{Martin~Camalich}},
  \bibnamefont{and} \bibinfo{author}{\bibfnamefont{J.~A.} \bibnamefont{Oller}},
  \bibinfo{journal}{Phys. Rev.} \textbf{\bibinfo{volume}{D85}},
  \bibinfo{pages}{051503} (\bibinfo{year}{2012}), \eprint{1110.3797}.

\bibitem[{\citenamefont{Ledwig et~al.}(2012)\citenamefont{Ledwig,
  Martin-Camalich, Pascalutsa, and Vanderhaeghen}}]{Ledwig:2011cx}
\bibinfo{author}{\bibfnamefont{T.}~\bibnamefont{Ledwig}},
  \bibinfo{author}{\bibfnamefont{J.}~\bibnamefont{Martin-Camalich}},
  \bibinfo{author}{\bibfnamefont{V.}~\bibnamefont{Pascalutsa}},
  \bibnamefont{and}
  \bibinfo{author}{\bibfnamefont{M.}~\bibnamefont{Vanderhaeghen}},
  \bibinfo{journal}{Phys. Rev.} \textbf{\bibinfo{volume}{D85}},
  \bibinfo{pages}{034013} (\bibinfo{year}{2012}), \eprint{1108.2523}.

\bibitem[{\citenamefont{Chen et~al.}(2013)\citenamefont{Chen, Yao, and
  Zheng}}]{Chen:2012nx}
\bibinfo{author}{\bibfnamefont{Y.-H.} \bibnamefont{Chen}},
  \bibinfo{author}{\bibfnamefont{D.-L.} \bibnamefont{Yao}}, \bibnamefont{and}
  \bibinfo{author}{\bibfnamefont{H.~Q.} \bibnamefont{Zheng}},
  \bibinfo{journal}{Phys. Rev.} \textbf{\bibinfo{volume}{D87}},
  \bibinfo{pages}{054019} (\bibinfo{year}{2013}), \eprint{1212.1893}.

\bibitem[{\citenamefont{Alvarez-Ruso et~al.}(2013)\citenamefont{Alvarez-Ruso,
  Ledwig, Martin~Camalich, and Vicente-Vacas}}]{Alvarez-Ruso:2013fza}
\bibinfo{author}{\bibfnamefont{L.}~\bibnamefont{Alvarez-Ruso}},
  \bibinfo{author}{\bibfnamefont{T.}~\bibnamefont{Ledwig}},
  \bibinfo{author}{\bibfnamefont{J.}~\bibnamefont{Martin~Camalich}},
  \bibnamefont{and} \bibinfo{author}{\bibfnamefont{M.~J.}
  \bibnamefont{Vicente-Vacas}}, \bibinfo{journal}{Phys. Rev.}
  \textbf{\bibinfo{volume}{D88}}, \bibinfo{pages}{054507}
  (\bibinfo{year}{2013}), \eprint{1304.0483}.

\bibitem[{\citenamefont{Ledwig et~al.}(2014)\citenamefont{Ledwig,
  Martin~Camalich, Geng, and Vicente~Vacas}}]{Ledwig:2014rfa}
\bibinfo{author}{\bibfnamefont{T.}~\bibnamefont{Ledwig}},
  \bibinfo{author}{\bibfnamefont{J.}~\bibnamefont{Martin~Camalich}},
  \bibinfo{author}{\bibfnamefont{L.~S.} \bibnamefont{Geng}}, \bibnamefont{and}
  \bibinfo{author}{\bibfnamefont{M.~J.} \bibnamefont{Vicente~Vacas}},
  \bibinfo{journal}{Phys. Rev.} \textbf{\bibinfo{volume}{D90}},
  \bibinfo{pages}{054502} (\bibinfo{year}{2014}), \eprint{1405.5456}.

\bibitem[{\citenamefont{Lensky et~al.}(2014)\citenamefont{Lensky, Alarcón, and
  Pascalutsa}}]{Lensky:2014dda}
\bibinfo{author}{\bibfnamefont{V.}~\bibnamefont{Lensky}},
  \bibinfo{author}{\bibfnamefont{J.~M.} \bibnamefont{Alarcón}},
  \bibnamefont{and}
  \bibinfo{author}{\bibfnamefont{V.}~\bibnamefont{Pascalutsa}},
  \bibinfo{journal}{Phys. Rev.} \textbf{\bibinfo{volume}{C90}},
  \bibinfo{pages}{055202} (\bibinfo{year}{2014}), \eprint{1407.2574}.

\bibitem[{\citenamefont{Yao et~al.}(2016)\citenamefont{Yao, Siemens, Bernard,
  Epelbaum, Gasparyan, Gegelia, Krebs, and Mei{\ss}ner}}]{Yao:2016vbz}
\bibinfo{author}{\bibfnamefont{D.-L.} \bibnamefont{Yao}},
  \bibinfo{author}{\bibfnamefont{D.}~\bibnamefont{Siemens}},
  \bibinfo{author}{\bibfnamefont{V.}~\bibnamefont{Bernard}},
  \bibinfo{author}{\bibfnamefont{E.}~\bibnamefont{Epelbaum}},
  \bibinfo{author}{\bibfnamefont{A.~M.} \bibnamefont{Gasparyan}},
  \bibinfo{author}{\bibfnamefont{J.}~\bibnamefont{Gegelia}},
  \bibinfo{author}{\bibfnamefont{H.}~\bibnamefont{Krebs}}, \bibnamefont{and}
  \bibinfo{author}{\bibfnamefont{U.-G.} \bibnamefont{Mei{\ss}ner}},
  \bibinfo{journal}{JHEP} \textbf{\bibinfo{volume}{05}}, \bibinfo{pages}{038}
  (\bibinfo{year}{2016}), \eprint{1603.03638}.

\bibitem[{\citenamefont{Siemens et~al.}(2016)\citenamefont{Siemens, Bernard,
  Epelbaum, Gasparyan, Krebs, and Mei{\ss}ner}}]{Siemens:2016hdi}
\bibinfo{author}{\bibfnamefont{D.}~\bibnamefont{Siemens}},
  \bibinfo{author}{\bibfnamefont{V.}~\bibnamefont{Bernard}},
  \bibinfo{author}{\bibfnamefont{E.}~\bibnamefont{Epelbaum}},
  \bibinfo{author}{\bibfnamefont{A.}~\bibnamefont{Gasparyan}},
  \bibinfo{author}{\bibfnamefont{H.}~\bibnamefont{Krebs}}, \bibnamefont{and}
  \bibinfo{author}{\bibfnamefont{U.-G.} \bibnamefont{Mei{\ss}ner}},
  \bibinfo{journal}{Phys. Rev.} \textbf{\bibinfo{volume}{C94}},
  \bibinfo{pages}{014620} (\bibinfo{year}{2016}), \eprint{1602.02640}.

\bibitem[{\citenamefont{Hiller~Blin
  et~al.}(2015{\natexlab{a}})\citenamefont{Hiller~Blin, Ledwig, and
  Vicente~Vacas}}]{Blin:2014rpa}
\bibinfo{author}{\bibfnamefont{A.~N.} \bibnamefont{Hiller~Blin}},
  \bibinfo{author}{\bibfnamefont{T.}~\bibnamefont{Ledwig}}, \bibnamefont{and}
  \bibinfo{author}{\bibfnamefont{M.~J.} \bibnamefont{Vicente~Vacas}},
  \bibinfo{journal}{Phys. Lett.} \textbf{\bibinfo{volume}{B747}},
  \bibinfo{pages}{217} (\bibinfo{year}{2015}{\natexlab{a}}),
  \eprint{1412.4083}.

\bibitem[{\citenamefont{Hiller~Blin et~al.}(2016)\citenamefont{Hiller~Blin,
  Ledwig, and Vicente~Vacas}}]{Blin:2016itn}
\bibinfo{author}{\bibfnamefont{A.~N.} \bibnamefont{Hiller~Blin}},
  \bibinfo{author}{\bibfnamefont{T.}~\bibnamefont{Ledwig}}, \bibnamefont{and}
  \bibinfo{author}{\bibfnamefont{M.~J.} \bibnamefont{Vicente~Vacas}},
  \bibinfo{journal}{Phys. Rev.} \textbf{\bibinfo{volume}{D93}},
  \bibinfo{pages}{094018} (\bibinfo{year}{2016}), \eprint{1602.08967}.

\bibitem[{\citenamefont{Hilt et~al.}(2013{\natexlab{b}})\citenamefont{Hilt,
  Lehnhart, Scherer, and Tiator}}]{Hilt:2013fda}
\bibinfo{author}{\bibfnamefont{M.}~\bibnamefont{Hilt}},
  \bibinfo{author}{\bibfnamefont{B.~C.} \bibnamefont{Lehnhart}},
  \bibinfo{author}{\bibfnamefont{S.}~\bibnamefont{Scherer}}, \bibnamefont{and}
  \bibinfo{author}{\bibfnamefont{L.}~\bibnamefont{Tiator}},
  \bibinfo{journal}{Phys. Rev.} \textbf{\bibinfo{volume}{C88}},
  \bibinfo{pages}{055207} (\bibinfo{year}{2013}{\natexlab{b}}),
  \eprint{1309.3385}.

\bibitem[{\citenamefont{Ericson and Weise}(1988)}]{Ericson:1988gk}
\bibinfo{author}{\bibfnamefont{T.~E.~O.} \bibnamefont{Ericson}}
  \bibnamefont{and} \bibinfo{author}{\bibfnamefont{W.}~\bibnamefont{Weise}},
  \emph{\bibinfo{title}{{Pions and Nuclei}}}, vol.~\bibinfo{volume}{74}
  (\bibinfo{publisher}{Clarendon Press}, \bibinfo{address}{Oxford, UK},
  \bibinfo{year}{1988}), ISBN \bibinfo{isbn}{0198520085},
  \urlprefix\url{http://www-spires.fnal.gov/spires/find/books/www?cl=QC793.5.M42E75::1988}.

\bibitem[{\citenamefont{Lensky and Pascalutsa}(2010)}]{Lensky:2009uv}
\bibinfo{author}{\bibfnamefont{V.}~\bibnamefont{Lensky}} \bibnamefont{and}
  \bibinfo{author}{\bibfnamefont{V.}~\bibnamefont{Pascalutsa}},
  \bibinfo{journal}{Eur. Phys. J.} \textbf{\bibinfo{volume}{C65}},
  \bibinfo{pages}{195} (\bibinfo{year}{2010}), \eprint{0907.0451}.

\bibitem[{\citenamefont{Hiller~Blin
  et~al.}(2015{\natexlab{b}})\citenamefont{Hiller~Blin, Gutsche, Ledwig, and
  Lyubovitskij}}]{Blin:2015era}
\bibinfo{author}{\bibfnamefont{A.}~\bibnamefont{Hiller~Blin}},
  \bibinfo{author}{\bibfnamefont{T.}~\bibnamefont{Gutsche}},
  \bibinfo{author}{\bibfnamefont{T.}~\bibnamefont{Ledwig}}, \bibnamefont{and}
  \bibinfo{author}{\bibfnamefont{V.~E.} \bibnamefont{Lyubovitskij}},
  \bibinfo{journal}{Phys. Rev.} \textbf{\bibinfo{volume}{D92}},
  \bibinfo{pages}{096004} (\bibinfo{year}{2015}{\natexlab{b}}),
  \eprint{1509.00955}.

\bibitem[{\citenamefont{Yao et~al.}(2018)\citenamefont{Yao, Alvarez-Ruso,
  Hiller~Blin, and Vicente~Vacas}}]{Yao:2018pzc}
\bibinfo{author}{\bibfnamefont{D.-L.} \bibnamefont{Yao}},
  \bibinfo{author}{\bibfnamefont{L.}~\bibnamefont{Alvarez-Ruso}},
  \bibinfo{author}{\bibfnamefont{A.~N.} \bibnamefont{Hiller~Blin}},
  \bibnamefont{and} \bibinfo{author}{\bibfnamefont{M.~J.}
  \bibnamefont{Vicente~Vacas}}, \bibinfo{journal}{Phys. Rev.}
  \textbf{\bibinfo{volume}{D98}}, \bibinfo{pages}{076004}
  (\bibinfo{year}{2018}), \eprint{1806.09364}.

\bibitem[{\citenamefont{Yao et~al.}(2019)\citenamefont{Yao, Alvarez-Ruso, and
  Vicente~Vacas}}]{Yao:2019avf}
\bibinfo{author}{\bibfnamefont{D.-L.} \bibnamefont{Yao}},
  \bibinfo{author}{\bibfnamefont{L.}~\bibnamefont{Alvarez-Ruso}},
  \bibnamefont{and} \bibinfo{author}{\bibfnamefont{M.~J.}
  \bibnamefont{Vicente~Vacas}}, \bibinfo{journal}{Phys. Lett.}
  \textbf{\bibinfo{volume}{B794}}, \bibinfo{pages}{109} (\bibinfo{year}{2019}),
  \eprint{1901.00773}.

\bibitem[{\citenamefont{Hemmert et~al.}(1997)\citenamefont{Hemmert, Holstein,
  and Kambor}}]{Hemmert:1996xg}
\bibinfo{author}{\bibfnamefont{T.~R.} \bibnamefont{Hemmert}},
  \bibinfo{author}{\bibfnamefont{B.~R.} \bibnamefont{Holstein}},
  \bibnamefont{and} \bibinfo{author}{\bibfnamefont{J.}~\bibnamefont{Kambor}},
  \bibinfo{journal}{Phys. Lett.} \textbf{\bibinfo{volume}{B395}},
  \bibinfo{pages}{89} (\bibinfo{year}{1997}), \eprint{hep-ph/9606456}.

\bibitem[{\citenamefont{Bernard et~al.}(1994)\citenamefont{Bernard, Kaiser, and
  Mei{\ss}ner}}]{Bernard:1993xh}
\bibinfo{author}{\bibfnamefont{V.}~\bibnamefont{Bernard}},
  \bibinfo{author}{\bibfnamefont{N.}~\bibnamefont{Kaiser}}, \bibnamefont{and}
  \bibinfo{author}{\bibfnamefont{U.~G.} \bibnamefont{Mei{\ss}ner}},
  \bibinfo{journal}{Phys. Lett.} \textbf{\bibinfo{volume}{B331}},
  \bibinfo{pages}{137} (\bibinfo{year}{1994}), \eprint{hep-ph/9312307}.

\bibitem[{\citenamefont{Alvarez-Ruso et~al.}(2018)}]{Alvarez-Ruso:2017oui}
\bibinfo{author}{\bibfnamefont{L.}~\bibnamefont{Alvarez-Ruso}}
  \bibnamefont{et~al.}, \bibinfo{journal}{Prog. Part. Nucl. Phys.}
  \textbf{\bibinfo{volume}{100}}, \bibinfo{pages}{1} (\bibinfo{year}{2018}),
  \eprint{1706.03621}.

\bibitem[{\citenamefont{Alvarez-Ruso et~al.}(2014)\citenamefont{Alvarez-Ruso,
  Hayato, and Nieves}}]{Alvarez-Ruso:2014bla}
\bibinfo{author}{\bibfnamefont{L.}~\bibnamefont{Alvarez-Ruso}},
  \bibinfo{author}{\bibfnamefont{Y.}~\bibnamefont{Hayato}}, \bibnamefont{and}
  \bibinfo{author}{\bibfnamefont{J.}~\bibnamefont{Nieves}},
  \bibinfo{journal}{New J. Phys.} \textbf{\bibinfo{volume}{16}},
  \bibinfo{pages}{075015} (\bibinfo{year}{2014}), \eprint{1403.2673}.

\bibitem[{\citenamefont{Chew et~al.}(1957)\citenamefont{Chew, Goldberger, Low,
  and Nambu}}]{Chew:1957tf}
\bibinfo{author}{\bibfnamefont{G.~F.} \bibnamefont{Chew}},
  \bibinfo{author}{\bibfnamefont{M.~L.} \bibnamefont{Goldberger}},
  \bibinfo{author}{\bibfnamefont{F.~E.} \bibnamefont{Low}}, \bibnamefont{and}
  \bibinfo{author}{\bibfnamefont{Y.}~\bibnamefont{Nambu}},
  \bibinfo{journal}{Phys. Rev.} \textbf{\bibinfo{volume}{106}},
  \bibinfo{pages}{1345} (\bibinfo{year}{1957}).

\bibitem[{\citenamefont{Sandorfi et~al.}(2011)\citenamefont{Sandorfi, Hoblit,
  Kamano, and Lee}}]{Sandorfi:2010uv}
\bibinfo{author}{\bibfnamefont{A.~M.} \bibnamefont{Sandorfi}},
  \bibinfo{author}{\bibfnamefont{S.}~\bibnamefont{Hoblit}},
  \bibinfo{author}{\bibfnamefont{H.}~\bibnamefont{Kamano}}, \bibnamefont{and}
  \bibinfo{author}{\bibfnamefont{T.~S.~H.} \bibnamefont{Lee}},
  \bibinfo{journal}{J. Phys.} \textbf{\bibinfo{volume}{G38}},
  \bibinfo{pages}{053001} (\bibinfo{year}{2011}), \eprint{1010.4555}.

\bibitem[{\citenamefont{Pascalutsa and Phillips}(2003)}]{Pascalutsa:2002pi}
\bibinfo{author}{\bibfnamefont{V.}~\bibnamefont{Pascalutsa}} \bibnamefont{and}
  \bibinfo{author}{\bibfnamefont{D.~R.} \bibnamefont{Phillips}},
  \bibinfo{journal}{Phys. Rev.} \textbf{\bibinfo{volume}{C67}},
  \bibinfo{pages}{055202} (\bibinfo{year}{2003}), \eprint{nucl-th/0212024}.

\bibitem[{\citenamefont{Fettes et~al.}(2000)\citenamefont{Fettes, Mei{\ss}ner,
  Mojzis, and Steininger}}]{Fettes:2000gb}
\bibinfo{author}{\bibfnamefont{N.}~\bibnamefont{Fettes}},
  \bibinfo{author}{\bibfnamefont{U.-G.} \bibnamefont{Mei{\ss}ner}},
  \bibinfo{author}{\bibfnamefont{M.}~\bibnamefont{Mojzis}}, \bibnamefont{and}
  \bibinfo{author}{\bibfnamefont{S.}~\bibnamefont{Steininger}},
  \bibinfo{journal}{Annals Phys.} \textbf{\bibinfo{volume}{283}},
  \bibinfo{pages}{273} (\bibinfo{year}{2000}), \bibinfo{note}{[Erratum: Annals
  Phys.288,249(2001)]}, \eprint{hep-ph/0001308}.

\bibitem[{\citenamefont{Pascalutsa}(2008)}]{Pascalutsa:2007yg}
\bibinfo{author}{\bibfnamefont{V.}~\bibnamefont{Pascalutsa}},
  \bibinfo{journal}{Prog. Part. Nucl. Phys.} \textbf{\bibinfo{volume}{61}},
  \bibinfo{pages}{27} (\bibinfo{year}{2008}), \eprint{0712.3919}.

\bibitem[{\citenamefont{Pascalutsa et~al.}(2007)\citenamefont{Pascalutsa,
  Vanderhaeghen, and Yang}}]{Pascalutsa:2006up}
\bibinfo{author}{\bibfnamefont{V.}~\bibnamefont{Pascalutsa}},
  \bibinfo{author}{\bibfnamefont{M.}~\bibnamefont{Vanderhaeghen}},
  \bibnamefont{and} \bibinfo{author}{\bibfnamefont{S.~N.} \bibnamefont{Yang}},
  \bibinfo{journal}{Phys. Rept.} \textbf{\bibinfo{volume}{437}},
  \bibinfo{pages}{125} (\bibinfo{year}{2007}), \eprint{hep-ph/0609004}.

\bibitem[{\citenamefont{Shtabovenko et~al.}(2016)\citenamefont{Shtabovenko,
  Mertig, and Orellana}}]{Shtabovenko:2016sxi}
\bibinfo{author}{\bibfnamefont{V.}~\bibnamefont{Shtabovenko}},
  \bibinfo{author}{\bibfnamefont{R.}~\bibnamefont{Mertig}}, \bibnamefont{and}
  \bibinfo{author}{\bibfnamefont{F.}~\bibnamefont{Orellana}},
  \bibinfo{journal}{Comput. Phys. Commun.} \textbf{\bibinfo{volume}{207}},
  \bibinfo{pages}{432} (\bibinfo{year}{2016}), \eprint{1601.01167}.

\bibitem[{\citenamefont{Mertig et~al.}(1991)\citenamefont{Mertig, Bohm, and
  Denner}}]{Mertig:1990an}
\bibinfo{author}{\bibfnamefont{R.}~\bibnamefont{Mertig}},
  \bibinfo{author}{\bibfnamefont{M.}~\bibnamefont{Bohm}}, \bibnamefont{and}
  \bibinfo{author}{\bibfnamefont{A.}~\bibnamefont{Denner}},
  \bibinfo{journal}{Comput. Phys. Commun.} \textbf{\bibinfo{volume}{64}},
  \bibinfo{pages}{345} (\bibinfo{year}{1991}).

\bibitem[{\citenamefont{Lehmann et~al.}(1955)\citenamefont{Lehmann, Symanzik,
  and Zimmermann}}]{Lehmann:1954rq}
\bibinfo{author}{\bibfnamefont{H.}~\bibnamefont{Lehmann}},
  \bibinfo{author}{\bibfnamefont{K.}~\bibnamefont{Symanzik}}, \bibnamefont{and}
  \bibinfo{author}{\bibfnamefont{W.}~\bibnamefont{Zimmermann}},
  \bibinfo{journal}{Nuovo Cim.} \textbf{\bibinfo{volume}{1}},
  \bibinfo{pages}{205} (\bibinfo{year}{1955}).

\bibitem[{\citenamefont{Fuchs et~al.}(1996)}]{Fuchs:1996ja}
\bibinfo{author}{\bibfnamefont{M.}~\bibnamefont{Fuchs}} \bibnamefont{et~al.},
  \bibinfo{journal}{Phys. Lett.} \textbf{\bibinfo{volume}{B368}},
  \bibinfo{pages}{20} (\bibinfo{year}{1996}).

\bibitem[{\citenamefont{Bergstrom et~al.}(1996)\citenamefont{Bergstrom, Vogt,
  Igarashi, Keeter, Hallin, Retzlaff, Skopik, and Booth}}]{Bergstrom:1996fq}
\bibinfo{author}{\bibfnamefont{J.~C.} \bibnamefont{Bergstrom}},
  \bibinfo{author}{\bibfnamefont{J.~M.} \bibnamefont{Vogt}},
  \bibinfo{author}{\bibfnamefont{R.}~\bibnamefont{Igarashi}},
  \bibinfo{author}{\bibfnamefont{K.~J.} \bibnamefont{Keeter}},
  \bibinfo{author}{\bibfnamefont{E.~L.} \bibnamefont{Hallin}},
  \bibinfo{author}{\bibfnamefont{G.~A.} \bibnamefont{Retzlaff}},
  \bibinfo{author}{\bibfnamefont{D.~M.} \bibnamefont{Skopik}},
  \bibnamefont{and} \bibinfo{author}{\bibfnamefont{E.~C.} \bibnamefont{Booth}},
  \bibinfo{journal}{Phys. Rev.} \textbf{\bibinfo{volume}{C53}},
  \bibinfo{pages}{R1052} (\bibinfo{year}{1996}).

\bibitem[{\citenamefont{Bergstrom et~al.}(1997)\citenamefont{Bergstrom,
  Igarashi, and Vogt}}]{Bergstrom:1997jc}
\bibinfo{author}{\bibfnamefont{J.~C.} \bibnamefont{Bergstrom}},
  \bibinfo{author}{\bibfnamefont{R.}~\bibnamefont{Igarashi}}, \bibnamefont{and}
  \bibinfo{author}{\bibfnamefont{J.~M.} \bibnamefont{Vogt}},
  \bibinfo{journal}{Phys. Rev.} \textbf{\bibinfo{volume}{C55}},
  \bibinfo{pages}{2016} (\bibinfo{year}{1997}).

\bibitem[{\citenamefont{Schmidt et~al.}(2001)}]{Schmidt:2001vg}
\bibinfo{author}{\bibfnamefont{A.}~\bibnamefont{Schmidt}} \bibnamefont{et~al.},
  \bibinfo{journal}{Phys. Rev. Lett.} \textbf{\bibinfo{volume}{87}},
  \bibinfo{pages}{232501} (\bibinfo{year}{2001}), \bibinfo{note}{[Erratum:
  Phys. Rev. Lett.110,039903(2013)]}, \eprint{nucl-ex/0105010}.

\bibitem[{\citenamefont{Blanpied et~al.}(2001)}]{Blanpied:2001ae}
\bibinfo{author}{\bibfnamefont{G.}~\bibnamefont{Blanpied}}
  \bibnamefont{et~al.}, \bibinfo{journal}{Phys. Rev.}
  \textbf{\bibinfo{volume}{C64}}, \bibinfo{pages}{025203}
  (\bibinfo{year}{2001}).

\bibitem[{\citenamefont{Rossi et~al.}(1973)}]{Rossi:1973wf}
\bibinfo{author}{\bibfnamefont{V.}~\bibnamefont{Rossi}} \bibnamefont{et~al.},
  \bibinfo{journal}{Nuovo Cim.} \textbf{\bibinfo{volume}{A13}},
  \bibinfo{pages}{59} (\bibinfo{year}{1973}).

\bibitem[{\citenamefont{Benz et~al.}(1973)}]{Benz:1974tt}
\bibinfo{author}{\bibfnamefont{P.}~\bibnamefont{Benz}} \bibnamefont{et~al.}
  (\bibinfo{collaboration}{Aachen-Bonn-Hamburg-Heidelberg-Muenchen}),
  \bibinfo{journal}{Nucl. Phys.} \textbf{\bibinfo{volume}{B65}},
  \bibinfo{pages}{158} (\bibinfo{year}{1973}).

\bibitem[{\citenamefont{Salomon et~al.}(1984)\citenamefont{Salomon, Measday,
  Poutissou, and Robertson}}]{Salomon:1983xn}
\bibinfo{author}{\bibfnamefont{M.}~\bibnamefont{Salomon}},
  \bibinfo{author}{\bibfnamefont{D.~F.} \bibnamefont{Measday}},
  \bibinfo{author}{\bibfnamefont{J.~M.} \bibnamefont{Poutissou}},
  \bibnamefont{and} \bibinfo{author}{\bibfnamefont{B.~C.}
  \bibnamefont{Robertson}}, \bibinfo{journal}{Nucl. Phys.}
  \textbf{\bibinfo{volume}{A414}}, \bibinfo{pages}{493} (\bibinfo{year}{1984}).

\bibitem[{\citenamefont{Bagheri et~al.}(1988)\citenamefont{Bagheri, Aniol,
  Entezami, Hasinoff, Measday, Poutissou, Salomon, and
  Robertson}}]{Bagheri:1987kf}
\bibinfo{author}{\bibfnamefont{A.}~\bibnamefont{Bagheri}},
  \bibinfo{author}{\bibfnamefont{K.~A.} \bibnamefont{Aniol}},
  \bibinfo{author}{\bibfnamefont{F.}~\bibnamefont{Entezami}},
  \bibinfo{author}{\bibfnamefont{M.~D.} \bibnamefont{Hasinoff}},
  \bibinfo{author}{\bibfnamefont{D.~F.} \bibnamefont{Measday}},
  \bibinfo{author}{\bibfnamefont{J.~M.} \bibnamefont{Poutissou}},
  \bibinfo{author}{\bibfnamefont{M.}~\bibnamefont{Salomon}}, \bibnamefont{and}
  \bibinfo{author}{\bibfnamefont{B.~C.} \bibnamefont{Robertson}},
  \bibinfo{journal}{Phys. Rev.} \textbf{\bibinfo{volume}{C38}},
  \bibinfo{pages}{875} (\bibinfo{year}{1988}).

\bibitem[{SAI()}]{SAID}
\emph{\bibinfo{title}{{INS Data Analysis Center}}},
  \urlprefix\url{http://gwdac.phys.gwu.edu/}.

\bibitem[{\citenamefont{Strandberg et~al.}(2018)}]{Strandberg:2018djk}
\bibinfo{author}{\bibfnamefont{B.}~\bibnamefont{Strandberg}}
  \bibnamefont{et~al.} (\bibinfo{year}{2018}), \eprint{1812.03023}.

\bibitem[{\citenamefont{Walker et~al.}(1963)\citenamefont{Walker, Palfrey,
  Haxby, and Nefkens}}]{Walker:1963zzb}
\bibinfo{author}{\bibfnamefont{R.~J.} \bibnamefont{Walker}},
  \bibinfo{author}{\bibfnamefont{T.~R.} \bibnamefont{Palfrey}},
  \bibinfo{author}{\bibfnamefont{R.~O.} \bibnamefont{Haxby}}, \bibnamefont{and}
  \bibinfo{author}{\bibfnamefont{B.~M.~K.} \bibnamefont{Nefkens}},
  \bibinfo{journal}{Phys. Rev.} \textbf{\bibinfo{volume}{132}},
  \bibinfo{pages}{2656} (\bibinfo{year}{1963}).

\bibitem[{\citenamefont{Fissum et~al.}(1996)\citenamefont{Fissum, Caplan,
  Hallin, Skopik, Vogt, Frodyma, Rosenzweig, Storm, O'Rielly, and
  Garrow}}]{Fissum:1996fi}
\bibinfo{author}{\bibfnamefont{K.~G.} \bibnamefont{Fissum}},
  \bibinfo{author}{\bibfnamefont{H.~S.} \bibnamefont{Caplan}},
  \bibinfo{author}{\bibfnamefont{E.~L.} \bibnamefont{Hallin}},
  \bibinfo{author}{\bibfnamefont{D.~M.} \bibnamefont{Skopik}},
  \bibinfo{author}{\bibfnamefont{J.~M.} \bibnamefont{Vogt}},
  \bibinfo{author}{\bibfnamefont{M.}~\bibnamefont{Frodyma}},
  \bibinfo{author}{\bibfnamefont{D.~P.} \bibnamefont{Rosenzweig}},
  \bibinfo{author}{\bibfnamefont{D.~W.} \bibnamefont{Storm}},
  \bibinfo{author}{\bibfnamefont{G.~V.} \bibnamefont{O'Rielly}},
  \bibnamefont{and} \bibinfo{author}{\bibfnamefont{K.~R.}
  \bibnamefont{Garrow}}, \bibinfo{journal}{Phys. Rev.}
  \textbf{\bibinfo{volume}{C53}}, \bibinfo{pages}{1278} (\bibinfo{year}{1996}).

\bibitem[{\citenamefont{Ahrens et~al.}(2004)}]{Ahrens:2004pf}
\bibinfo{author}{\bibfnamefont{J.}~\bibnamefont{Ahrens}} \bibnamefont{et~al.}
  (\bibinfo{collaboration}{GDH, A2}), \bibinfo{journal}{Eur. Phys. J.}
  \textbf{\bibinfo{volume}{A21}}, \bibinfo{pages}{323} (\bibinfo{year}{2004}).

\bibitem[{\citenamefont{Bauer et~al.}(2012)\citenamefont{Bauer, Bernauer, and
  Scherer}}]{Bauer:2012pv}
\bibinfo{author}{\bibfnamefont{T.}~\bibnamefont{Bauer}},
  \bibinfo{author}{\bibfnamefont{J.~C.} \bibnamefont{Bernauer}},
  \bibnamefont{and} \bibinfo{author}{\bibfnamefont{S.}~\bibnamefont{Scherer}},
  \bibinfo{journal}{Phys. Rev.} \textbf{\bibinfo{volume}{C86}},
  \bibinfo{pages}{065206} (\bibinfo{year}{2012}), \eprint{1209.3872}.

\bibitem[{\citenamefont{Patrignani et~al.}(2016)}]{Patrignani:2016xqp}
\bibinfo{author}{\bibfnamefont{C.}~\bibnamefont{Patrignani}}
  \bibnamefont{et~al.} (\bibinfo{collaboration}{Particle Data Group}),
  \bibinfo{journal}{Chin. Phys.} \textbf{\bibinfo{volume}{C40}},
  \bibinfo{pages}{100001} (\bibinfo{year}{2016}).

\bibitem[{\citenamefont{Yao et~al.}(2017)\citenamefont{Yao, Alvarez-Ruso, and
  Vicente-Vacas}}]{Yao:2017fym}
\bibinfo{author}{\bibfnamefont{D.-L.} \bibnamefont{Yao}},
  \bibinfo{author}{\bibfnamefont{L.}~\bibnamefont{Alvarez-Ruso}},
  \bibnamefont{and} \bibinfo{author}{\bibfnamefont{M.~J.}
  \bibnamefont{Vicente-Vacas}}, \bibinfo{journal}{Phys. Rev.}
  \textbf{\bibinfo{volume}{D96}}, \bibinfo{pages}{116022}
  (\bibinfo{year}{2017}), \eprint{1708.08776}.

\bibitem[{\citenamefont{Bernard et~al.}(2013)\citenamefont{Bernard, Epelbaum,
  Krebs, and Mei{\ss}ner}}]{Bernard:2012hb}
\bibinfo{author}{\bibfnamefont{V.}~\bibnamefont{Bernard}},
  \bibinfo{author}{\bibfnamefont{E.}~\bibnamefont{Epelbaum}},
  \bibinfo{author}{\bibfnamefont{H.}~\bibnamefont{Krebs}}, \bibnamefont{and}
  \bibinfo{author}{\bibfnamefont{U.-G.} \bibnamefont{Mei{\ss}ner}},
  \bibinfo{journal}{Phys. Rev.} \textbf{\bibinfo{volume}{D87}},
  \bibinfo{pages}{054032} (\bibinfo{year}{2013}), \eprint{1209.2523}.

\bibitem[{\citenamefont{Epelbaum et~al.}(2015)\citenamefont{Epelbaum, Krebs,
  and Meißner}}]{Epelbaum:2014efa}
\bibinfo{author}{\bibfnamefont{E.}~\bibnamefont{Epelbaum}},
  \bibinfo{author}{\bibfnamefont{H.}~\bibnamefont{Krebs}}, \bibnamefont{and}
  \bibinfo{author}{\bibfnamefont{U.~G.} \bibnamefont{Meißner}},
  \bibinfo{journal}{Eur. Phys. J.} \textbf{\bibinfo{volume}{A51}},
  \bibinfo{pages}{53} (\bibinfo{year}{2015}), \eprint{1412.0142}.

\bibitem[{\citenamefont{Schumann et~al.}(2010)}]{Schumann:2010js}
\bibinfo{author}{\bibfnamefont{S.}~\bibnamefont{Schumann}}
  \bibnamefont{et~al.}, \bibinfo{journal}{Eur. Phys. J.}
  \textbf{\bibinfo{volume}{A43}}, \bibinfo{pages}{269} (\bibinfo{year}{2010}),
  \eprint{1001.3626}.

\bibitem[{\citenamefont{Korkmaz et~al.}(1999)}]{Korkmaz:1999sg}
\bibinfo{author}{\bibfnamefont{E.}~\bibnamefont{Korkmaz}} \bibnamefont{et~al.},
  \bibinfo{journal}{Phys. Rev. Lett.} \textbf{\bibinfo{volume}{83}},
  \bibinfo{pages}{3609} (\bibinfo{year}{1999}).

\bibitem[{\citenamefont{McPherson et~al.}(1964)\citenamefont{McPherson, Gates,
  Kenney, and Swanson}}]{McPherson:1964zz}
\bibinfo{author}{\bibfnamefont{D.~A.} \bibnamefont{McPherson}},
  \bibinfo{author}{\bibfnamefont{D.~C.} \bibnamefont{Gates}},
  \bibinfo{author}{\bibfnamefont{R.~W.} \bibnamefont{Kenney}},
  \bibnamefont{and} \bibinfo{author}{\bibfnamefont{W.~P.}
  \bibnamefont{Swanson}}, \bibinfo{journal}{Phys. Rev.}
  \textbf{\bibinfo{volume}{136}}, \bibinfo{pages}{B1465}
  (\bibinfo{year}{1964}).

\bibitem[{\citenamefont{White et~al.}(1960)\citenamefont{White, Schectman, and
  Chasan}}]{White:1960ukk}
\bibinfo{author}{\bibfnamefont{D.~H.} \bibnamefont{White}},
  \bibinfo{author}{\bibfnamefont{R.~M.} \bibnamefont{Schectman}},
  \bibnamefont{and} \bibinfo{author}{\bibfnamefont{B.~M.}
  \bibnamefont{Chasan}}, \bibinfo{journal}{Phys. Rev.}
  \textbf{\bibinfo{volume}{120}}, \bibinfo{pages}{614} (\bibinfo{year}{1960}).

\bibitem[{\citenamefont{Wang}(1992)}]{WA92}
\bibinfo{author}{\bibfnamefont{M.}~\bibnamefont{Wang}}, \bibinfo{type}{{Ph. D.}
  thesis}, \bibinfo{school}{University of Kentucky} (\bibinfo{year}{1992}).

\bibitem[{\citenamefont{Liu}(1994)}]{LI94}
\bibinfo{author}{\bibfnamefont{K.}~\bibnamefont{Liu}}, \bibinfo{type}{{Ph. D.}
  thesis}, \bibinfo{school}{University of Kentucky} (\bibinfo{year}{1994}).

\bibitem[{\citenamefont{Gegelia et~al.}(2016)\citenamefont{Gegelia,
  Mei{\ss}ner, Siemens, and Yao}}]{Gegelia:2016pjm}
\bibinfo{author}{\bibfnamefont{J.}~\bibnamefont{Gegelia}},
  \bibinfo{author}{\bibfnamefont{U.-G.} \bibnamefont{Mei{\ss}ner}},
  \bibinfo{author}{\bibfnamefont{D.}~\bibnamefont{Siemens}}, \bibnamefont{and}
  \bibinfo{author}{\bibfnamefont{D.-L.} \bibnamefont{Yao}},
  \bibinfo{journal}{Phys. Lett.} \textbf{\bibinfo{volume}{B763}},
  \bibinfo{pages}{1} (\bibinfo{year}{2016}), \eprint{1608.00517}.

\end{thebibliography}

\end{document}